\documentclass[onecolumn,authoryear]{els-mrw} 

\usepackage{amsmath,amssymb,amsfonts,amsthm,makeidx,graphicx}
\usepackage{txfonts}
\usepackage{helvet}
\usepackage[dvipsnames]{xcolor}
\usepackage{hyperref}
\hypersetup{
    colorlinks=true,
    linkcolor=blue,
    filecolor=magenta,      
    urlcolor=RubineRed,
    pdftitle={Overleaf Example},
    pdfpagemode=FullScreen,
    }
\newcommand{\ch}[1]{\textcolor{black}{#1}}

\newcommand{\js}{} 

\newcommand\aj{\js{AJ}}
\newcommand\araa{\js{ARA\&A}}
\newcommand\apj{\js{ApJ}}
\newcommand\apjl{\js{ApJL}}     
\newcommand\apjs{\js{ApJS}}
\newcommand\aap{\js{A\&A}}
\newcommand\aapr{\js{A\&A~Rv}}
\newcommand\aaps{\js{A\&AS}}
\newcommand\mnras{\js{MNRAS}}
\newcommand\pasp{\js{PASP}}
\newcommand\nat{\js{Nature}}
\newcommand\na{\js{NewA}}
\newcommand\nar{\js{NewAR}}
\newcommand\pasa{\js{PASA}}
\newcommand\rmxaa{\js{RMxAA}}
\newcommand\fcp{\js{FCP}}
\newcommand\physrep{\js{PhR}}
\newcommand\prd{\js{PhysRevD}}
\newcommand\apss{\js{ApSS}}

\begin{document}

\chapter{The Spectral Energy Distributions of Galaxies}\label{chap1}

\author[1]{Kartheik G. Iyer}%
\author[2]{Camilla Pacifici}%
\author[3]{Gabriela Calistro-Rivera}%
\author[4]{Christopher C. Lovell}%

\address[1]{\orgname{Columbia University}, \orgdiv{Columbia Astrophysics Lab}, \orgaddress{550 W 120th St, New York, NY 10010, USA.}}
\address[2]{\orgname{Space Telescope Science Institute}, \orgaddress{3700 San Martin Drive, Baltimore, MD 21218, USA.}}
\address[3]{\orgname{European Southern Observatory (ESO)}, \orgaddress{Karl-Schwarzschild-Straße 2, 85748 Garching bei München, Germany}}
\address[4]{\orgname{Institute of Cosmology and Gravitation}, \orgaddress{University of Portsmouth, Burnaby Road, Portsmouth PO1 3FX, UK}}

\articletag{Chapter Article tagline: update of previous edition,, reprint..}

\maketitle

\begin{glossary}[Glossary]

\term{Active Galactic Nucleus}: The bright central region of a galaxy powered by a massive black hole actively consuming surrounding material.

\term{Attenuation}: The \ch{wavelength-dependent} reduction in the observed brightness of galactic light due to absorption and scattering by dust, distinct from extinction as it accounts for the complex geometry of stars and dust in galaxies.

\term{Composite Stellar Population}: The combined population of stars with different ages and chemical compositions in a galaxy.

\term{Dust}: Small solid particles (typically 5-250 nm) in the ISM that absorb ultraviolet and optical light and re-emit it at infrared wavelengths.

\term{Initial Mass Function}: A distribution describing how many stars of each mass are created in a given stellar population.

\term{Metallicity}: The abundance of elements heavier than hydrogen and helium in a galaxy's stars and gas, indicating its chemical composition.

\term{Nebular Emission}: Light produced by ionized gas in galaxies, often appearing as emission lines in galaxy spectra and tracing recent star formation.

\term{Photometric Redshift}: A galaxy's redshift estimated using broadband photometry rather than spectroscopy, providing a less precise but more efficient measure of galaxy distances.

\term{Redshift}: The stretching of light to longer wavelengths due to the expansion of the universe, used to measure cosmic distances and ages.

\term{Simple Stellar Population}: A group of stars formed at the same time with the same chemical composition, representing the basic building block for modeling galaxy spectra.

\term{Spectral Energy Distribution}: A measure of galaxy brightness (energy or flux density) as a function of frequency or wavelength.

\term{Spectroscopic Redshift}: A precise measurement of a galaxy's redshift obtained by identifying \ch{discrete, wavelength-specific} spectral features.

\term{Star Formation History}: The record of a galaxy's star formation rate over time, encoded in its present-day stellar populations.

\term{Star Formation Rate}: The rate at which a galaxy is currently converting gas into new stars, typically measured in solar masses per year.

\term{Stellar Population Synthesis}: The technique of modeling galaxy spectra by combining the light from stellar populations of different ages and chemical compositions.

\end{glossary}

\begin{abstract}[Abstract]
The spectral energy distribution (SED) of a galaxy represents the distribution of electromagnetic radiation emitted across all wavelengths, from radio waves to gamma rays. The galaxy SED is akin to its fingerprint, and serves as a fundamental tool in modern astrophysics. It enables researchers to determine crucial properties of galaxies, including their star-formation rates, stellar populations, dust content, and evolutionary state. By analyzing galactic SEDs, astronomers can reconstruct the physical processes occurring within galaxies and trace their evolutionary histories. This article explores our current understanding of the components that contribute to galactic SEDs, the observational techniques used to measure them, and their applications in understanding galaxy formation and evolution. 
\end{abstract}

\begin{BoxTypeA}[chap1:box1]{Key points}
\begin{itemize}
    \item A galaxy's \textbf{spectral energy distribution (SED)} is composed of light from its component stars, ionized gas, dust, and active galactic nuclei. It contains several signatures of its past formation and current physical state. 
    \item One of the primary uses of a distant galaxy's SED is to determine its distance from us, using photometric or spectroscopic redshift techniques. 
    \item  Modeling an SED involves combining models for emission from these different sources, whose contributions vary according to the wavelength range being studied. Stellar light tends to dominate the UV-to-NIR portion of the rest-frame spectrum, AGN and dust dominate the mid-IR, dust and gas dominate the far-IR, and AGN and X-ray binaries tend to dominate the X-ray and radio regimes. 
    \item SED fitting is used to refer to a set of techniques used to infer the physical properties of a galaxy by modeling its SED, which include its stellar mass, star formation rate and history, chemical abundance, ionization state, amount and nature of dust, supermassive black hole properties, and more. 
    \item Observational SEDs have evolved tremendously over last few decades, with upcoming surveys set to obtain spectra and photometry for hundreds of millions of galaxies with observatories like Rubin, Roman, and Euclid. 
\end{itemize}
\end{BoxTypeA}

\begin{glossary}[Nomenclature]
\begin{tabular}{@{}lp{34pc}@{}@{}lp{34pc}@{}}
AGN &Active Galactic Nucleus \\
CSP &Composite Stellar Population\\
IFU &Integral field Unit\\
IMF &Initial Mass Function\\ 
IR &Infrared (also near-IR (NIR), mid-IR and far-IR)\\
ISM &Interstellar Medium\\
PAH &Polycyclic Aromatic Hydrocarbon\\
PSF &Point Spread Function\\
SED &Spectral Energy Distribution\\
SFH &Star Formation History\\
SFR &Star Formation Rate\\
SMBH &Supermassive Black Hole\\
SNR &Signal-to-Noise Ratio\\
SPS &Stellar Population Synthesis\\
SSP &Simple Stellar Population\\
UV &Ultraviolet\\
z &redshift (photo-z and spec-z) \\
$\tau$ &optical depth\\
$A_V$ &extinction in the V-band\\
$\beta$ &UV spectral slope\\
$\chi^2$ &chi-squared statistic\\
$F_\lambda$ &Flux Density per unit wavelength\\
$F_\nu$ &Flux Density per unit frequency\\
$M_*$ &Stellar mass\\
$Z$ &Metallicity\\
\end{tabular}
\end{glossary}

\section{Introduction}
\label{introduction}

The spectral energy distribution (SED) of a galaxy represents one of the most powerful tools we have for understanding the universe beyond our own galaxy. Similar to how the rings of a tree tell the story of its life, a galaxy's SED - a measure of its flux density across the electromagnetic spectrum - provides a distinctive signature that encodes crucial information about its nature and history. The galaxy SED allows astronomers to determine fundamental properties including the galaxy's star formation activity, stellar populations, dust content, and evolutionary state.

Our ability to decode these cosmic fingerprints has evolved dramatically over time. Beginning with pioneering work on modeling stellar populations in the 1970s \citep{1968ApJ...151..547T, 1976ApJ...203...52T, 1980FCPh....5..287T, 1973ApJ...179..731F}, astronomers have steadily built up sophisticated frameworks for interpreting different components of galaxy SEDs - from the ultraviolet light of young stars to the infrared glow of heated dust. Modern telescopes, ranging from the Hubble Space Telescope (HST) to the Atacama Large Millimeter Array (ALMA), have revolutionized our capabilities by providing unprecedented views across the electromagnetic spectrum. The launch of the James Webb Space Telescope (JWST) has opened new windows into galaxy evolution during the universe's earliest epochs.

Today, SED analysis forms the foundation of modern extragalactic astronomy and cosmology. By studying galaxy SEDs, we can effectively use telescopes as time machines - examining snapshots of galaxies at different cosmic epochs to reconstruct how they have grown and evolved over billions of years. This has profound implications for our understanding of fundamental questions in astronomy: \textit{How do galaxies form and evolve? What drives their star formation and quenching? How do supermassive black holes influence their host galaxies?} As we enter an era of increasingly powerful survey telescopes and sophisticated analysis techniques, galaxy SEDs continue to offer critical insights into these essential questions about our cosmic history.

This article explores the key components that contribute to galaxy SEDs (Section \ref{sec:sedmodel}), the observational techniques used to measure them (Section \ref{sec:sedprop}), and the methods astronomers use to extract physical insights from these cosmic tree rings (Section \ref{sec:fitting}). We aim to provide both an overview of this fundamental tool in modern astrophysics and a glimpse into how it continues to advance our understanding of the universe.

\section{Galaxy Observations}

\begin{figure}
    \centering
    \includegraphics[width=\linewidth, page=1, trim={0 33cm 0 0},clip]{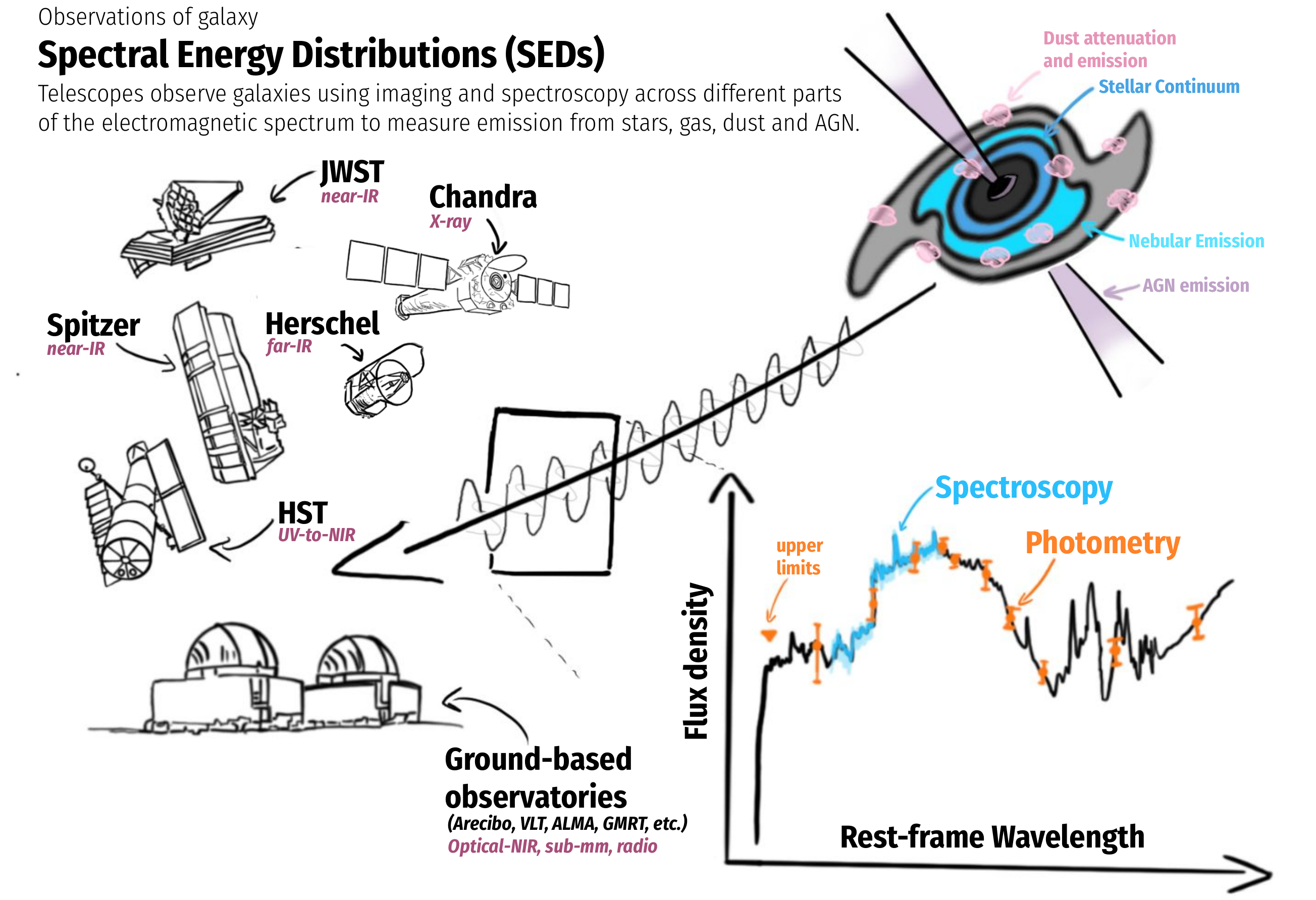}
    \caption{The SED of a galaxy measures its flux density as a function of frequency or wavelength. A variety of ground- and space-based telescopes are used to observe the light from galaxies across a wide range of wavelengths and redshifts. Astronomers use the compiled spectroscopic and photometric data to infer the physical properties of galaxies and study their evolution.}
    \label{fig:sed_observations}
\end{figure}

Galaxy observations across a range of wavelengths (Figure \ref{fig:sed_observations}) form the foundation of modern extragalactic astronomy and cosmology. The process of studying galaxies involves a series of steps: observing with telescopes; reduction of the raw data; object detection; data extraction; and catalog building, followed by the measurement of their physical properties.
In this section, we will not discuss the observations and data reduction (which can be found in standard texts, e.g. \citealt{burns2021practical}), focusing instead on the different ways of measuring galaxy SEDs, and galaxy surveys based on these measurements.

\subsection{Photometry}

Photometry is the measurement of the brightness of astronomical sources through a photometric bandpass, also called filters. This can be thought of in a way similar to the RGB filters or UV filters available for modern cameras or the solar filters used to view eclipses. These filters, when used in a telescope in conjunction with a charge coupled device (CCD) or calorimeter, allow astronomers to measure the brightness or flux density of an object in a fixed wavelength range. Measurements using photometric filters across a range of wavelengths create a coarse sampling of a galaxy's spectrum, and subsequently \ch{allow us to} estimate their distances and physical properties. Important concepts when dealing with photometry are the field of view, the spatial resolution, and the filter coverage.

The field of view refers to the total area visible in an image (the wider the area, the more objects can be detected at the same time), while the spatial or angular resolution indicates the level of detail within that view, meaning how well fine features can be distinguished within the image. Generally, increasing the field of view results in a decrease in spatial resolution, as the same number of pixels are spread over a larger area. However, modern techniques such as adaptive optics and exquisite sensors provide large a field-of-view (FoV) without compromising on spatial resolution.

Broad wavelength coverage can be obtained with wide-band filters (about 2000 to 10,000 \AA{} width) to \ch{sample the average SED over a wide wavelength range}, medium-band filters (about 500 to 2000\AA{} width) to identify specific features like a break in the SED, or narrow-band filters (about 30 to 500 \AA{} width) to target \ch{emission lines and absorption features}. For galaxies, photometry is typically performed in multiple bands spanning the ultraviolet, optical, and infrared regions of the SED. Common photometric systems used in galaxy studies are the Johnson-Cousins system (\textit{UBVRI}) and the Sloan Digital Sky Survey (SDSS) system (\textit{ugriz}). More recent surveys like \ch{UltraVISTA \citep{2012A&A...544A.156M} and PAU \citep{2009ApJ...691..241B} have shown the importance of medium- and narrow-band filters}, especially to measure \ch{galaxy redshifts} (see Section \ref{sec:redshift}). It is important to keep in mind that the narrower the filter, the longer the exposure time required to reach the necessary signal-to-noise ratio.

Photometry 
comes primarily in two flavors: aperture photometry and profile fitting. Aperture photometry involves measuring the total flux within a fixed circular or elliptical aperture centered on the source. This method is simple and computationally efficient but can be affected by contamination from nearby sources or background variations. Profile fitting, on the other hand, involves modeling the light distribution of the galaxy using a parametric function, such as the Sersic profile. This method is more robust to contamination and can provide additional information about the galaxy's morphology, but it is computationally more expensive and requires careful modeling. Some public codes used to perform photometry include SExtractor \citep{1996A&AS..117..393B}, photutils \citep{2016ascl.soft09011B}, TFIT \citep{2007PASP..119.1325L}, T-PHOT \citep{2015A&A...582A..15M}, PetroFit \citep{2022AJ....163..202G} and crowdsource \citep{2018ApJS..234...39S, 2021ascl.soft06004S}. 

\subsection{Spectroscopy}

Compared to photometry, spectroscopy uses a prism or grating to split the light from a galaxy into its component wavelengths and generally provides much more detailed measurements of its SED. Spectroscopic observations are often characterized by their slit or fiber configuration (i.e., how many objects can be observed at the same time), the wavelength range, and the spectral resolution. 
Slits and fibers collect the light from a small targeted region in the sky. The light is then dispersed through a grating, and collected on the detector. Spectroscopy allows for the identification of specific spectral features that can be used to accurately measure the properties \ch{and distances} of galaxies.

Galaxies can be observed through slits, in slitless mode, or through integral field unit (IFU) observations. Instruments can provide the option to use a single slit or to observe a number of objects at the same time with multi-object spectrographs. In slitless mode (often called \textit{grism}), the light from the whole field of view passes through a grating, thus dispersing the light of every object along a chosen orientation. This mode is extremely powerful in terms of discovery space, but can be challenging when fields are especially crowded. IFU spectroscopy involves a range of techniques used to collect spectra along a two-dimensional field-of-view, producing datacubes of spectra with fluxes in (x,y,$\lambda$) space\footnote{A summary of different IFU techniques can be found at: http://ifs.wikidot.com/.}. 

Most ground-based spectroscopy is limited to optical or radio frequencies because of difficulties in observing at other frequencies due to the Earth's atmospheric absorption. Space telescopes have played a significant part in observing galaxies in the UV-to-NIR, with spectra covering the observed wavelength range from $\sim 1500$\AA{} to $27\mu m$ accessible through HST and JWST. Rest-frame optical wavelengths are typically targeted to study absorption features that probe the ages and metallicities of the stars in galaxies, and emission lines to study the properties of the ionized gas within galaxies. 

The spectral resolution is dictated by the choice of grating and ranges from low ($\lambda/\Delta\lambda=R=100$) to medium ($R=1000$ to $2000$) to high (typically $R=2000$ to $10,000$, but can go as high as 100,000 with the ELT/TMT). Low resolution is cheaper (in terms of exposure time) than high resolution, but comes at the expense of the level of details in the spectra. Low resolution spectroscopy is generally used for large samples to measure the strongest features like optical emission lines or continuum for bright galaxies. Medium resolution spectroscopy (at high signal to noise) is sufficient to detect the different gas elements \ch{in emission} and optical absorption lines to constrain the star formation histories of galaxies. High resolution spectroscopy is generally used to measure the kinematics of the gas and the stars (when the exposure time is sufficient to detect the stellar continuum).

\subsection{Galaxy surveys}

Galaxy surveys play a crucial role in building our current understanding of galaxy evolution by providing large, systematic datasets across different wavelengths and cosmic epochs. They act as a census of galaxy populations, allowing us to understand how galaxies evolve over time both individually and \ch{at a population level}. Over the last few decades, there have been three main improvements in the data quality from galaxy surveys, which has been accompanied by corresponding advances in our techniques to analyze them by modeling and fitting SEDs. These are:  
\begin{enumerate}
    \item \textbf{The increase in wavelength baselines}: Synergies between different telescopes have enabled pan-chromatic surveys with large wavelength baselines. Since probing a galaxy's SED over a wider baseline can capture contributions from different sources, it helps break degeneracies between factors like dust and stellar populations that have traditionally plagued SED fitting.
    \item \textbf{Increased SNR}: Every newer generation of telescopes observes the sky with increased sensitivity. That, combined with deep- and ultra-deep surveys have obtained measurements of galaxy SEDs with unprecedented sensitivity across a wide range of epochs. The better SNR allows for SED fitting to distinguish between factors that affect galaxy SEDs at a few percent level, and constrain properties like stellar populations that were inaccessible with observations in the past.
    \item \textbf{Increased resolution with spectrophotometry}: Spectra, by dint of requiring more telescope time to observe a galaxy at a comparable depth to imaging, are generally available for a much smaller number of galaxies. However, large spectroscopic surveys, especially those that use integral field units to get a spectrum at every part of a galaxy, have increased the number of galaxies with spectra by orders of magnitude. Primarily, this helps determine the accuracy of redshifts and galaxy distances, but detailed spectra of the galaxy continuum help get better estimates of chemical enrichment, ISM conditions, AGN activity, and winds and outflows.
\end{enumerate}

\subsubsection{Surveys by wavelength}
Galaxy surveys can be primarily organized by the intended wavelength range, since this often determines both the intended science case and the instruments used to observe galaxies. These can be roughly categorized into:
\begin{itemize}
    \item \textbf{Optical surveys:} Optical surveys measure the rest-UV (at high-z) and optical portions of galaxy SEDs, and are used to characterize their stellar masses and star formation activity, stellar populations, and chemical and dust content. The Sloan Digital Sky Survey (SDSS; \citealt{2000AJ....120.1579Y}) has been a cornerstone of this field through its systematic five-band \textit{(ugriz}) photometry and spectroscopy of over a million galaxies. Its multiple phases have had different scientific priorities and have provided increasingly detailed spectroscopic coverage and photometric depth across 18 data releases\footnote{https://www.sdss.org/dr18/  and https://skyserver.sdss.org/edr/en/sdss/skyserver/ - SDSS was one of the first astronomical projects to budget for storage and pipeline software in addition to operating costs.}. The full list of optical surveys is too large to list here, but a few other notable ones include 2dF\footnote{http://www.2dfgrs.net/} \citep{2001MNRAS.328.1039C},  2MASS \citep{2006AJ....131.1163S}, PanSTARRS \citep{2002SPIE.4836..154K, 2016arXiv161205560C}, GAMA\footnote{https://www.gama-survey.org/} 
\citep{2011MNRAS.413..971D}, CANDELS\footnote{https://www.ipac.caltech.edu/project/candels} \citep{2011ApJS..197...35G, 2011ApJS..197...36K},  COSMOS\footnote{https://cosmos.astro.caltech.edu/page/hst} \citep{2007ApJS..172....1S, 2022ApJS..258...11W}, and DES\footnote{https://www.darkenergysurvey.org/} \citep{2005astro.ph.10346T}  and DESI \citep{2016arXiv161100036D, 2024arXiv240919066S}, which use both ground- and space-based facilities to obtain both photometric observations and spectra. SDSS spectroscopic surveys have evolved from simple redshift surveys to detailed studies of galaxy properties with integral field spectroscopy through MaNGA. High-redshift spectroscopic surveys like VANDELS and LEGA-C have provided crucial calibration data for SED fitting at earlier cosmic epochs.
\item \textbf{Ultraviolet Surveys:} Ultraviolet surveys using instruments like GALEX \citep{2014AdSpR..53..900B, 2017ApJS..230...24B}, Swift, ASTROSAT \citep{2017AJ....154..128T, 2021ApJ...919..101P} and HST \citep{2015AJ....149...51C}  and UVEX \citep{2021arXiv211115608K} measure the rest-UV galaxy SEDs, and are used primarily to determine UV-based star formation rates. Unlike optical telescopes which can operate from the ground, most UV observatories need to be in space due to the Earth's atmosphere. 
\item \textbf{(near-to-mid) Infrared surveys:} Infrared surveys probe the rest-IR, optical and sometimes even UV portion of galaxy SEDs depending on the redshift, and are used for a host of different science-cases. The rest-frame near- to mid-IR wavelengths are sensitive to old stars (and are extremely important for robustly constraining stellar mass, dust emission and AGN torus emission, see Section \ref{sec:sedprop}), and are accessible primarily through space-based observatories. IR surveys are also very useful for detecting galaxies that are so heavily dust-obscured that they are not visible in the UV or optical. Prior to space-based telescopes, numerous ground-based observatories measured galaxy SEDs up to the K-band ($\sim 2.2 \mu$m; e.g. \citealt{1994MNRAS.266...65G}).  Ultra-deep surveys like the Hubble Deep Fields (HDF, HUDF) and their successors have pushed our understanding of galaxy SEDs to the highest redshifts. These pencil-beam surveys provide extremely deep multiwavelength coverage, allowing for the study of galaxy evolution across cosmic time. While many surveys with HST and Spitzer like GOALS \citep{2009PASP..121..559A}, SINGS \citep{2003PASP..115..928K}, SEDs \citep{2013ApJ...769...80A} and the UKIRT Infrared Deep Sky Survey (UKIDSS; \citealt{2007MNRAS.379.1599L}) and its successor, the VISTA surveys \citep{2004Msngr.117...27E} (including UltraVISTA \citep{2012A&A...544A.156M}), VIDEO: VISTA Deep Extragalactic Observations, SWIRE: Spitzer Wide-area InfraRed Extragalactic Survey, SIMPLE: Spitzer IRAC/MUSYC Pub-
lic Legacy in E-CDFS, SpUDS: Spitzer UKIDSS Ultra Deep Survey\footnote{The line between optical and IR surveys is blurred as surveys collate data across multiple instruments to create multiwavelength surveys. Surveys like CANDELS and COSMOS extend well into the infrared by combining HST and Spitzer observations in surveys like S-CANDELS and S-COSMOS.}, have probed galaxies to $z\sim 6-8$, the launch of JWST has spurred an entirely new regime of observations with surveys like CEERS, \ch{NGDEEP, PRIMER, JADES, CANUCS, UNCOVER, GLASS-JWST} and more. 
\item \textbf{Far-Infrared to sub-mm surveys:} In contrast to NIR, the far-IR SEDs of galaxies probe the dust mass and temperature \ch{(in the overall ISM as well as the central regions surrounding the AGN)}, star formation and molecular gas content (\ch{the latter traced through CO lines}; see review by \citealt{2014ARA&A..52..373L}). Using observatories like ALMA, NOEMA, and SMA on the ground and Herschel in space, some notable surveys \ch{in this wavelength regime} include ASPECS (searching for CO and [CII] emission in the HUDF), ALPINE (focusing on [CII] in $z\sim 4-6$ galaxies),  A3COSMOS (collecting all public ALMA data), GOODS-ALMA, ALMAQuest, PHIBSS,  SCUBA-2, HerMES, and H-ATLAS, \ch{and CO Surveys including} xCOLD GASS (Extended CO Legacy Database for GASS) and PHIBSS: IRAM Plateau de Bure HIgh-z Blue Sequence Survey using IRAM. 
\item \textbf{Radio continuum and HI surveys:} Radio continuum surveys are a useful tracer of unobscured star formation and AGN activity, while surveys measuring the strength of the 21cm line emission probe ionized gas in galaxies. Some notable radio surveys include VLA-COSMOS \citep{2017A&A...602A...1S}, FIRST \citep{1997ApJ...475..479W}, VLASS \citep[Very Large Array Sky Survey][]{2020PASP..132c5001L}, EMU \citep[Evolutionary Map of the Universe with ASKAP][]{2011PASA...28..215N}, LoTSS \citep[LOFAR Two-metre Sky Survey][]{2017A&A...598A.104S}. Notable HI Surveys include ALFALFA \citep[Arecibo Legacy Fast ALFA survey][]{2005AJ....130.2598G}; WALLABY \citep[Widefield ASKAP L-band Legacy All-sky Blind surveY][]{2020Ap&SS.365..118K} and XGASS \citep[Extended GALEX Arecibo SDSS Survey][]{2018MNRAS.476..875C}. 
\item \textbf{X-ray surveys:} X-ray surveys use telescopes like Chandra and XMM-Newton to extend SED wavelength baselines and characterize AGN activity. Notable surveys include the Chandra Deep Fields \citep[CDF-N, CDF-S][]{2000ApJ...541...49H, 2004ApJS..155..271S};  XMM-XXL \citep{2016A&A...592A...1P} and eRASS \citep[eROSITA All-Sky Survey][]{2024A&A...685A.106B}.
\end{itemize}
\vspace{0.2cm}
In addition to the above, it is important to mention multiwavelength surveys that include datasets from more than one instrument in order to probe a broader wavelength baseline. A good example of this is the COSMOS field, where the original 2 deg$^2$ field has become one of the most extensively studied regions of the sky, with data spanning from X-ray to radio wavelengths through multiple targeted surveys. Complementary to this, spectrophotometric surveys combine the broad baseline from photometry with the spectral features and redshift precision measured with spectroscopy. A good example of this are follow-up programs to \ch{SDSS such as the Galex-SDSS-Wise Legacy Catalog (GSWLC; \citealt{2016ApJS..227....2S})}, or ALMAQuest \citep{2019ApJ...884L..33L}, which combines ALMA observations in the sub-mm with optical IFU spectroscopy from SDSS-IV MaNGA. 
These surveys, particularly when combined across different wavelengths, have been essential for developing and validating SED fitting techniques. They provide both the training sets for empirical methods and the validation data for testing theoretical models of galaxy evolution\footnote{One such effort in recent times is the Multimodal Universe collaboration \citep{}}.
These surveys, particularly when combined across different wavelengths, have been essential for developing and validating SED fitting techniques. They provide both the training sets for empirical methods and the validation data for testing theoretical models of galaxy evolution\footnote{One such effort in recent times is the Multimodal Universe collaboration \citep{}}.
These surveys, particularly when combined across different wavelengths, have been essential for developing and validating SED fitting techniques. They provide both the training sets for empirical methods and the validation data for testing theoretical models of galaxy evolution\footnote{One such effort to collect datasets across a range of facilities in recent times is the Multimodal Universe collaboration \citep{2024arXiv241202527T}.}.

\textbf{The next generation:} The next generation of surveys promises to revolutionize our understanding of galaxy SEDs through both ground and space-based facilities. 
From space, the Nancy Grace Roman Space Telescope will provide HST-quality imaging over much wider fields, while the Euclid mission will combine optical and near-infrared imaging with slitless spectroscopy across large areas of sky.
On the ground, the Vera C. Rubin Observatory's Legacy Survey of Space and Time (LSST) will provide unprecedented photometric depth in six bands (\textit{ugrizy}) over the entire southern sky, complemented by massive spectroscopic surveys from facilities like the Dark Energy Spectroscopic Instrument (DESI), the Prime Focus Spectrograph (PFS) on Subaru, and the 4m multi-object Spectroscopic Telescope (4MOST). Additional facilities like SPHEREx and the upcoming extremely large telescopes (ELTs) will further expand on depth, angular resolution, and survey area, and inform the next generation of data-driven studies of galaxy evolution.

\section{The anatomy of a galaxy's SED}
\label{sec:sedmodel}

\begin{figure}
    \centering
    \includegraphics[width=\linewidth, page=2, trim={0 15.9cm 0 0},clip]{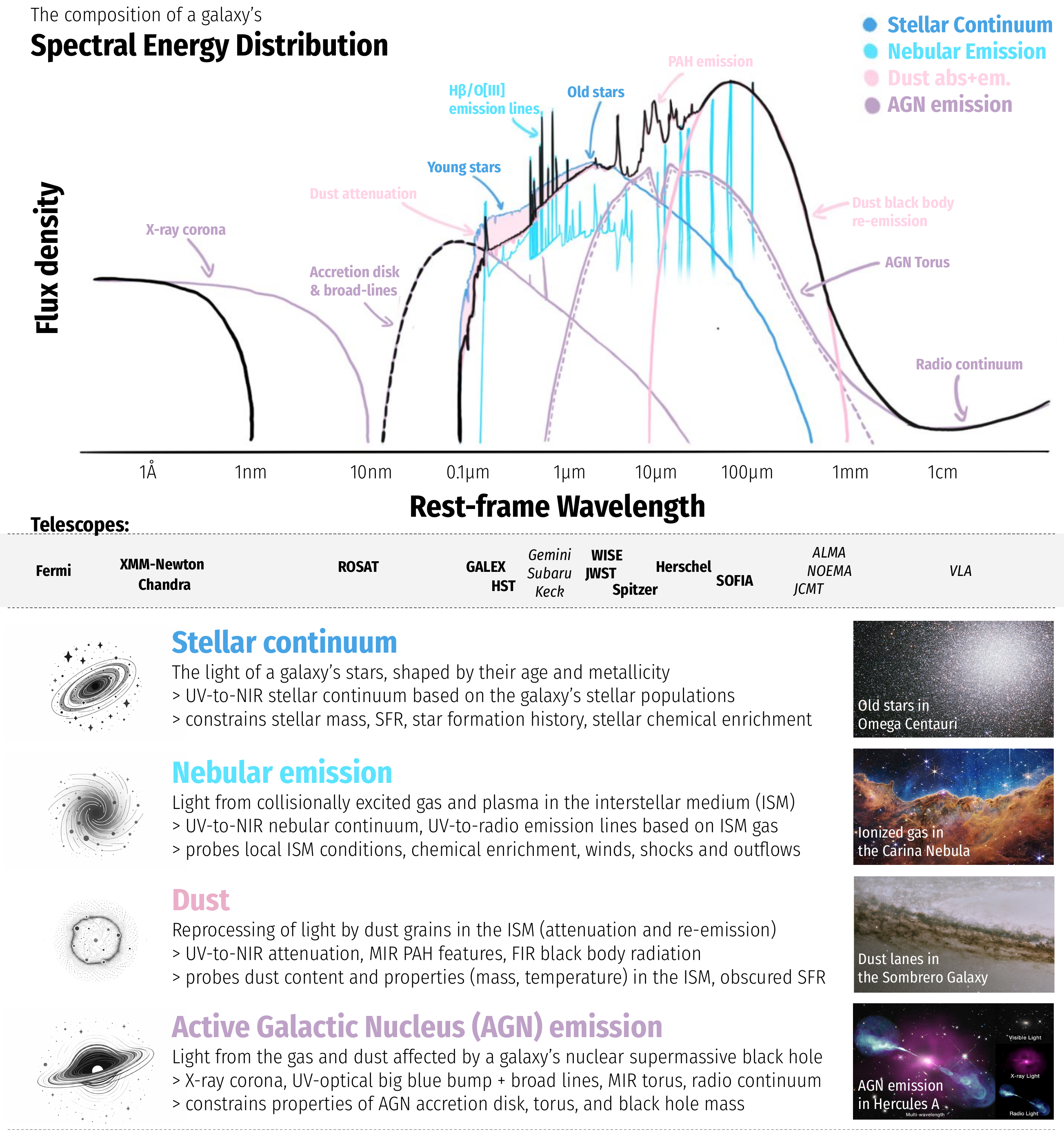}
    \caption{\textbf{Top:} A schematic of a galaxy's pan-chromatic SED, broken down by source of emission at different wavelengths. 
    \textbf{Bottom:} image credits (from top to bottom): Carina Nebula (NASA, ESA, CSA, and STScI), Omega Centauri (ESO), Sombrero Galaxy (NASA and the Hubble Heritage Team, STScI/AURA), and Hercules A (X-ray: NASA/CXC/SAO; visual: NASA/STScI; radio: NSF/NRAO/VLA).
    }
    \label{fig:sed_anatomy_main}
\end{figure}

A galaxy's SED encodes information about both its current physical state and past star formation activity, and contains contributions from various sources (shown in Figure \ref{fig:sed_anatomy_main}).  
Understanding these components and their contributions across different parts of the electromagnetic spectrum is crucial for interpreting galaxy SEDs and extracting meaningful physical properties. 
In this section, we will explore the primary components that make up a galaxy SED: the stellar continuum, the gas emission, the dust attenuation and re-emission, and active galactic nuclei (AGN). In this section, we will focus on how we can model each component of a galaxy's SED, while Section \ref{sec:sedprop} will delve into how we can infer physical properties from a galaxy's integrated SED.

\subsection{Stellar Continuum}

\begin{figure}
    \centering
    \includegraphics[width=0.9\linewidth, page=3, trim={0 4.2cm 0 0},clip]{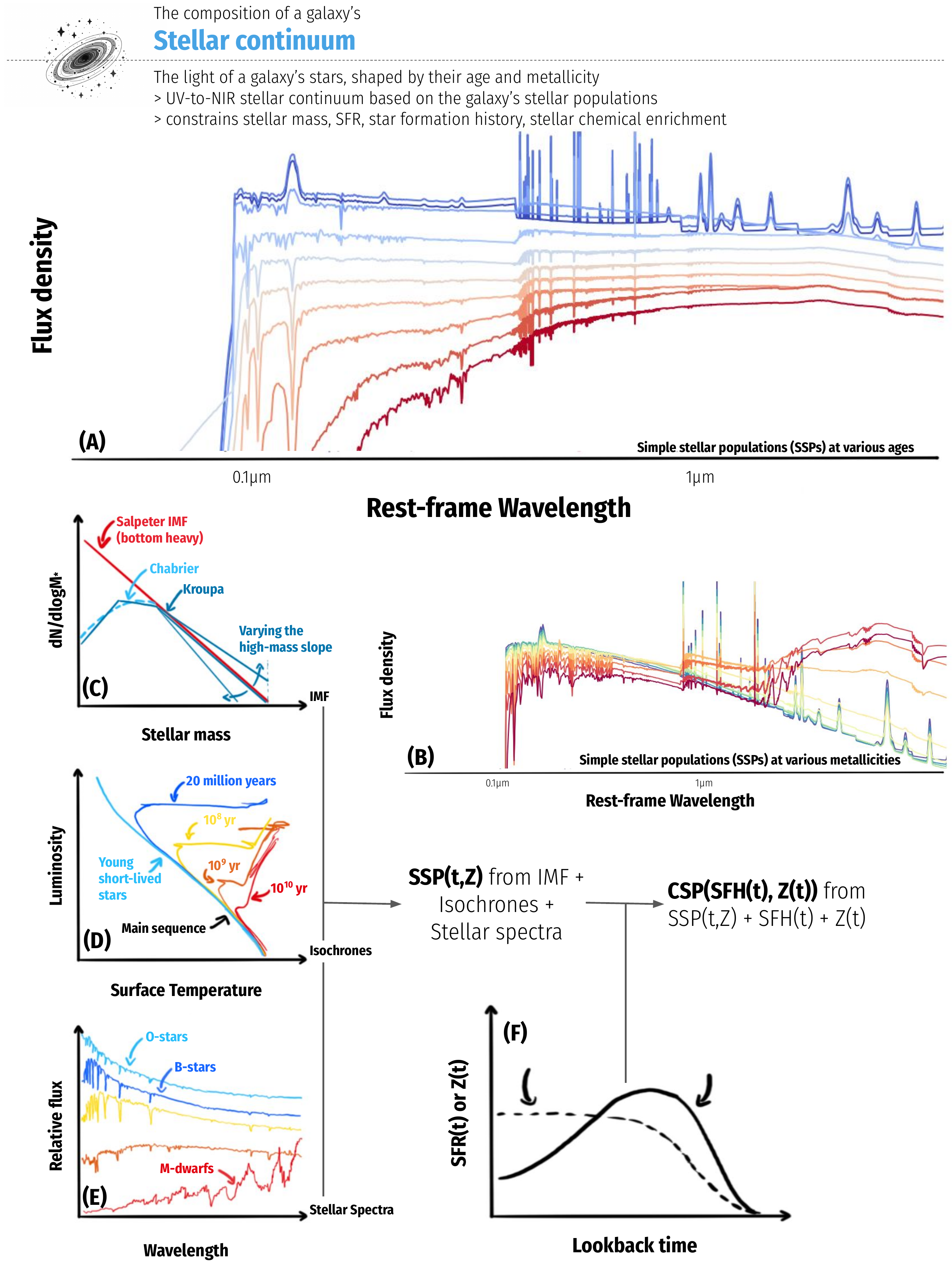}
    \caption{Creating a spectrum for the galaxy's stellar continuum. \textbf{(A)} Change in the relative flux of a galaxy's SED with varying age (blue showing the youngest to red showing the oldest) at fixed metallicity. 
    \textbf{(B)} Similar plot showing the effect of varying the metallicity at fixed age.
    The initial mass function \textbf{(C)} showing the distribution of stars as a function of mass,
    combined with stellar isochrones \textbf{(D)}, 
    and the spectra for different stellar types \textbf{(E)} can be used to generate the spectrum for a population of stars at a given age and metallicity.
    \textbf{(F)} The sum of SSPs weighted  by the galaxy's star formation and chemical enrichment history gives its stellar continuum.
    }
    \label{fig:sed_anatomy_stellar_continuum}
\end{figure}

The stellar continuum is the foundation of a galaxy SED, representing the combined emission from the galaxy's stellar populations. The shape of the stellar continuum depends on the properties of those stars, particularly their age and metallicity. Stellar Population Synthesis (SPS) can be used to generate the spectrum of a simple stellar population (SSP); when weighted by the star formation history (SFH), which describes the star formation rate at a given age, these models can be used to predict the composite stellar continuum of an entire galaxy. These SPS models are shown in Figure \ref{fig:sed_anatomy_stellar_continuum} and described briefly below, with a fuller treatment available in the review by \citealt{2013ARA&A..51..393C}.

\subsubsection{The Initial Mass Function}
 
The initial mass function (IMF) describes the distribution of stellar masses for a newly formed stellar population. Common choices for the IMF include Salpeter \citep{1955ApJ...121..161S}, Chabrier \citep{2003PASP..115..763C}, and Kroupa \citep{2001MNRAS.322..231K}\footnote{While all three IMF forms are similar at high-masses, the Salpeter IMF is generally more bottom-heavy, i.e. it contains more stars at the low-mass end. To convert a stellar mass estimated using a Salpeter IMF to Chabrier or Kroupa one can multiply by a factor of 1.67 or 1.53 respectively \citep{2011ApJ...737...67M, 2014ApJS..214...15S}.}.  The choice of IMF can have a significant impact on the shape of the stellar continuum, as well as the inferred star formation rates and stellar masses. While the IMF is generally assumed to be universal\footnote{in part due to how difficult it is to obtain observational constraints for the IMF, especially its low-mass slope.}, measurements of the IMF at the cores of low-redshift elliptical galaxies are found to contain more low-mass stars than the canonical IMF \citep{2010ApJ...709.1195T, 2022ApJ...932..103G}, while measurements of the nebular emission in high-redshift galaxies indicate a top-heavy IMF, with more massive stars than expected \citep{2024MNRAS.534..523C}. For more information about the physical mechanisms responsible for the IMF, see the review by \cite{2024ARA&A..62...63H}.

\subsubsection{Stellar Population Synthesis}

The first step in modeling the stellar continuum is to get the spectrum of a population of galaxies at a single epoch or age. 
Stellar population synthesis models, such as those by Bruzual \& Charlot \citep{2003MNRAS.344.1000B}, PARSEC \citep{2012MNRAS.427..127B}, MILES \citep{2010MNRAS.404.1639V, 2011A&A...532A..95F}, MIST \citep{2016ApJ...823..102C}, BPASS \citep{2017PASA...34...58E} and $\alpha$-MC \citep{2024arXiv241021375P} provide templates for the SEDs of SSPs with different ages and metallicities. These models are often based on a mixture of theoretical and empirical stellar tracks and isochrones. 

Stellar tracks quantify the evolution of individual stars at different initial masses in terms of temperature, luminosity, and other properties such as chemical composition. They trace a star's life from the main sequence through various evolutionary phases. Isochrones are curves connecting points representing stars of different masses but the same age. They are created by combining multiple stellar tracks and show the expected distribution of stars at a given moment in time. In combination with an initial mass function (IMF), which describes the relative distribution of stars of different masses, it is possible to combine the spectra of stars of different ages to create a simple stellar population (SSP), which is the spectrum of a stellar population with a single age and metallicity. While SPS models are the foundation of most studies of the stellar continuum, they still contain many nuances in their creation and calibration, such as accounting for stellar rotation \citep{2024A&A...685A..40S, 2025arXiv250103133N}, or contributions from binary stellar populations and thermally pulsating asymptotic giant branch (TP-AGB) stars \citep{2006ApJ...652...85M}. For more detailed reviews, see \cite{2012ARA&A..50..251I, 2013ARA&A..51..393C, 2022ARA&A..60..455E}.

\ch{A growing body of literature shows that most stars tend to have a companion \citep[for a detailed treatment, see][]{2023ASPC..534..275O}. Observationally, this tends to produce strong signatures in the stellar continuum, especially at young ages since the lifetimes of massive stars are often extended through processes such as mass transfer and tidal interactions, leading to bluer colors and greater luminosities \citep{2006ApJ...641..252D, 2009MNRAS.396..276Z}. This has led to the development of dedicated SPS models that include binary evolution such as BPASS \citep{2017PASA...34...58E, 2018MNRAS.479...75S, 2025MNRAS.tmp..163B} and POSYDON \citep{2023ApJS..264...45F, 2024arXiv241102376A}. The inclusion of these binary templates in SED fitting can lead to changes in the inferred physical parameters of galaxies, and an increasing number of modern studies are both accounting for binary populations in SED fitting \citep{2022MNRAS.514.5706J, 2024arXiv240303908H}, and determining portions of galaxy SEDs (e.g., the rest-UV; \citealt{2013ApJ...776...37L} or the He II 1640\AA{} emission line \citealt{2020MNRAS.495.4605S}) that provide sensitive constraints on the binary fraction\footnote{Though these are much more difficult to measure if metallicity, dust extinction and degeneracies in the SFHs are not properly resolved.}. This is especially true at higher redshifts, where the contribution from binaries is expected to increase given their younger, metal-poor stellar populations.}

\subsubsection{Star Formation and Chemical Enrichment Histories}

A galaxy's star formation history (SFH) describes how its star formation rate has evolved over time as a result of factors including gas accretion, feedback from galactic winds and AGN, mergers and baryon cycling \citep{2020MNRAS.498..430I}. In conjunction with the choice of ch{SPS} models, the shape of the galaxy's SFH has the most effect on the resulting stellar spectrum, and directly determines related quantities such as its resulting stellar mass, mass- and light-weighted ages and star formation rate. \ch{Spectra of the stellar continuum are generally computed by combining SSPs weighted by a galaxy's star formation and chemical enrichment histories to create composite stellar populations (CSPs; \citealt{2004ApJ...612..168P})}.

Since young, massive stars tend to dominate the rest-UV end of the spectrum while older, lower-mass stars peak in the NIR, the imprints of a galaxy's SFH are present in the stellar continuum portion of the SED. However, since the sensitivity of stellar populations falls off with increasing age and young hot stars tend to dominate the spectrum, the amount of information about the SFH that can be recovered from a galaxy's spectrum is often limited to a small set of numbers, and is much more sensitive to younger stellar populations. The SFHs of galaxies from nearby observations, where we can count individual stars and thus reconstruct the history using color-magnitude diagrams (CMDs), and galaxies in numerical simulations indicate considerable short-timescale stochasticity, sometimes referred to as `burstiness'.

\textbf{Chemical enrichment history:} In addition to the SFH, a galaxy's stellar continuum and several spectral features, such as the strength of the Balmer break and the equivalent width of several absorption lines, is dependent on the metallicity history of the galaxy. The chemical enrichment of a galaxy is determined by the evolution of the different stellar populations with time \citep{1979ApJ...229.1046T, 1995ApJ...454...69P}, which produce elements heavier than Hydrogen and Helium through nucleosynthesis along different channels \citep[stellar evolution, supernovae, AGB stars etc.; see reviews by][]{2020ARA&A..58..363P, 2020ARA&A..58..727J}. As metal-rich stars explode as supernovae and drive interstellar winds, they also enrich the interstellar medium (ISM) of the galaxy, where the heavier elements can influence how efficiently the gas can cool and collapse to form the next generation of stars. This is further complicated by the inflow of pristine gas from cosmic filaments, the mixing of stars due to galactic dynamics, and the accumulation of ex-situ stars from mergers. The chemical enrichment history of a galaxy determines the metallicity of its stellar populations, which in turn influences the shape of the stellar continuum. Galaxies with higher metallicities tend to have redder stellar continua due to increased line blanketing and opacity.

\textbf{Alpha Enhancement:} Alpha enhancement refers to the relative abundance of elements formed through the $\alpha$-process (e.g., O, Mg, Si, Ca) compared to iron ($[\alpha/\mathrm{Fe}]$).
$\alpha$ elements are typically formed through Type II SNe, the end point of massive star evolution, whereas Iron peak elements are typically formed through Type Ia SNe.

Since these SNe occur on different timescales, the alpha enhancement is also sensitive to the underlying star formation history. In the SED, $\alpha$-enhancement is observable through slightly bluer colors and stronger absorption features. \ch{These features are also expected in early galaxies because core-collapse supernovae (which produce alpha elements) occur soon after star formation, while Type Ia supernovae (which produce iron-peak elements) require longer timescales \citep{2020ApJ...900..179K,2022MNRAS.512.5329B}.}

\textbf{Population III stars:} Believed to be the universe's first generation of stars \citep[see reviews by][]{2004ARA&A..42...79B, 2023ARA&A..61...65K}, SEDs of galaxies containing Pop III stars are much sought after using telescopes like JWST \citep{2011ApJ...740...13Z, 2023MNRAS.525.5328T, 2024A&A...687A..67M} . These stars are thought to form from pristine, metal-free gas and were likely extremely massive ($>100$M$_\odot$), resulting in much hotter surface temperatures ($>10^5$K) compared to later generations. Their metal-free atmospheres would produce a harder ionizing spectrum with stronger UV emission, and their SEDs would lack metal absorption features. While Population III stars have not been directly observed, their signatures might be detectable in very high redshift galaxies through their distinctive nebular emission features, including strong HeII emission at 1640\AA, which requires the hard ionizing spectrum characteristic of metal-free stars\footnote{Dedicated SPS models are being developed to predict their unique emission characteristics \citep[e.g. Yggdrasil,][]{2011ApJ...740...13Z, 2017ApJ...836...78Z}}.

\subsection{The Interstellar Medium}

\begin{figure}
    \centering
    \includegraphics[width=0.99\linewidth, page=4, trim={0 16.4cm 0 0},clip]{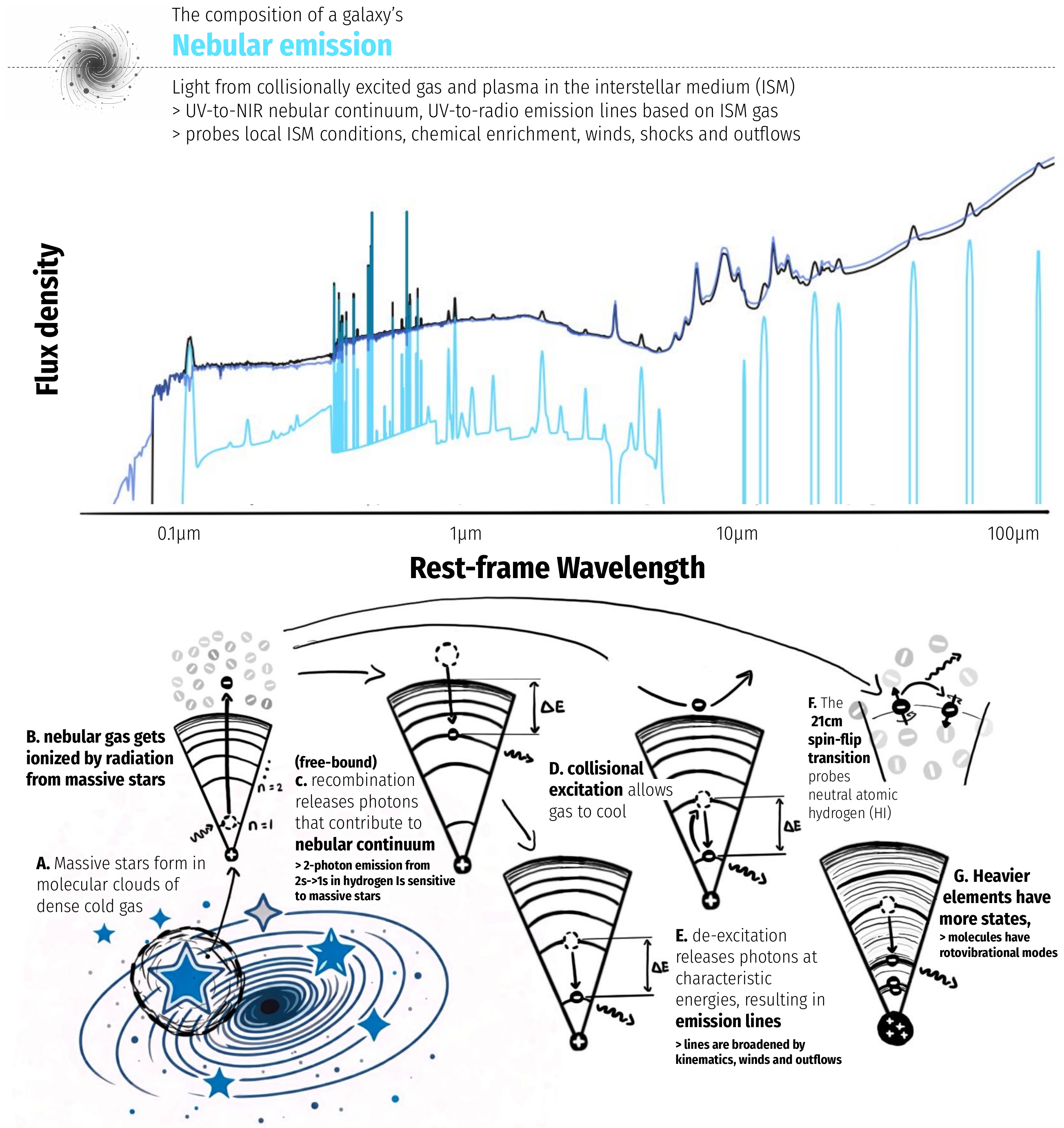}
    \caption{Nebular emission (light blue spectrum in top panel) provides a window to the physical properties of the gas in the interstellar medium. (bottom panels A-G) Schematics showing the ionization of ISM gas by massive stars and the resulting sources of emission.}
    \label{fig:sed_anatomy_gas}
\end{figure}

The interstellar medium (ISM) of galaxies is composed of multiphase gas (Table \ref{tab:ISMphases}), plasma and dust grains, which act as the reservoirs of fuel for star formation and are in turn affected by feedback from stars and AGN. In this section, we describe the emission from the gas and plasma, and discuss dust in Section \ref{sec:dust}, since it tends to produce distinct effects in the galaxy's SED. 

The gas content of galaxies, both atomic and molecular, contributes to their SEDs through collisionally excited and auroral lines, as well as rotovibrational transitions and spin-flip transitions. These are often grouped together under the term ``nebular emission'', referring to nebulae of ionized gases surrounding young stars that tend to produce such emission. The observable features of this type of emission provide valuable insights into the state of the ISM excited by the radiation of young stars in their birth-clouds and affected by shocks and outflows due to galactic winds and AGN feedback.
Interstellar gas can also lead to narrow absorption features in the continuum emission from the background stars.
These also provide key insights into the state of this intervening gas, such as its composition and ionization state. 

The nebular emission of galaxies varies with the chemical abundance and composition of the gas, the strength of the ionization potential, and the gas-to-dust ratio.
This section briefly describes the phases of the ISM gas, focusing on the spectral features of the ionized component which is strongest at optical wavelengths, and at how the emission can be modeled to derive important physical parameters. Further information regarding the physical mechanisms of star formation can be found in reviews by \cite{2007ARA&A..45..565M, 2024ARA&A..62..369S}, and reviews on the interstellar medium and nebular emission by \cite{2019ARA&A..57..511K, 2022ARA&A..60..319S, 2020ARA&A..58..157T}.

\subsubsection{The ISM gas of galaxies}

\begin{table}[]
    \centering
    \begin{tabular}{l c c c}
    \hline
    \hline
    Description & Phase & T(K) & n(cm$^{-3}$)\\
    \hline
    Cold molecular & H$_2$ & $\sim$15 & $>$100\\
    Cool atomic & HI & $\sim$80 & $\sim$30\\
    Warm atomic & HI & 10$^4$ & $\sim$0.5\\
    Diffuse ionized & HII & 10$^5$ & 10$^{-2}$\\
    Hot ionized & HII region & 10$^4$ & 10$^{3}$-10$^{4}$\\
    \hline
    \end{tabular}
    \caption{Phases, average temperatures, and densities of the ISM in galaxies.}
    \label{tab:ISMphases}
\end{table}

\noindent \textbf{Molecular gas - H$_2$}: Molecular gas is cold and dense and primarily composed of H$_2$, and is the primary fuel for star formation. CO molecules are exited by collisions with H$_2$ and emit at radio frequencies, and are easier to observe than H$_2$ itself \citep{2013ARA&A..51..207B}.

\noindent \textbf{Atomic gas - HI}: Atomic gas is primarily composed of neutral hydrogen (HI) and is the \textit{reservoir} of fuel that cools and gets converted to H2 to form stars. In diffuse interstellar clouds, neutral hydrogen is cool in the inner parts and warm in the outer parts, which can be partially ionized by the hard ultraviolet photons of the interstellar radiation field. Along with hydrogen, neutral heavy elements can also be found in diffuse clouds, especially oxygen and carbon, which are important cooling sources. Neutral hydrogen emits at a wavelength of 21-cm, which corresponds to the spin-flip transition between the two hyperfine levels of the fundamental state of HI\footnote{i.e. the change in the spin configuration of the electron and proton from parallel to antiparallel.}.

\noindent \textbf{Ionized gas}: Ionized gas can be found either in the form of diffuse intercloud gas or in the form of HII regions. The diffuse component is particularly hot, but it has extremely low density. As suggested by \cite{1977ApJ...218..148M}, the state of the interstellar medium is likely to be regulated by supernova explosions, which sweep out the gas, disrupting existing clouds and maintaining the intercloud medium in a hot ionized state. When gas in molecular clouds subsequently cools and star formation is triggered, the newly formed O- and B-type stars emit photons that ionize the birth cloud. This ionized gas cloud is called an HII region. HII regions can be easily observed detecting strong hydrogen and oxygen emission lines at optical wavelengths.

\subsubsection{Spectral features of the ionized gas in galaxies at UV-optical wavelengths}

The spectrum of an HII region is characterized by strong emission lines and a continuum component (see e.g., the models by \citealt{2017ApJ...840...44B} and \citealt{2021ApJ...922..170B}), coming from two sources: recombination processes and the radiative cooling of collisionally excited gas. 

\textbf{The continuum} consists of free–free (Bremsstrahlung radiation resulting from electron-ion interactions), free-bound (arising from radiative recombination events), and two-photon emission (from the decay of the metastable 2$^2S_{1/2}$ state of Hydrogen). This emission is the strongest at high ionization parameters (U; defined as the ratio of ionizing photon density to hydrogen density) and low metallicities. It is especially prominent near the Balmer break ($\sim 3646$\AA) in local young star clusters or in high redshift galaxies that are still in low metallicity environment.

\textbf{The emission-line} component is the result of recombination of ions and radiative de-excitation of collisionally excited molecules, atoms, and ions in the gas cloud \citep{Osterbrock2006, 2019ARA&A..57..511K}. If the cloud is optically thin, all photons issued from recombination can escape the nebula (case-A recombination). Otherwise, all photons issued from recombination that cascade to the ground state are immediately reabsorbed by a neutral hydrogen atom, so that all downward transitions to the n=1 level can be ignored (case-B recombination, more common in astrophysical contexts). In this case, every recombination must eventually yield a Balmer photon (transition to the n=2 level) at optical wavelengths. 
In case-B recombination, the luminosity ratio between the H$\alpha$ line (transition n = 3 $\rightarrow$ 2) at 6563\AA{} and the H$\beta$ line (n = 4 $\rightarrow$ 2) at 4861\AA{} is 2.87 for a nebula of temperature T = 10,000 K. The variation of this ratio, called Balmer decrement, is widely used to quantify the presence of dust in the ionized clouds.

The other main source of emission-line radiation in HII regions is the radiative de-excitation of collisionally excited ions, such as O+ , O++ , N+ and S+ . Some strong optical transitions are shown in Figure \ref{fig:sed_anatomy_gas}. The most prominent ones are [OII]3727, [OIII]5007, [NII]6548, [NII]6584 and [SII]6716, 6731, \ch{often called collisionally excited lines (CELs). Since the cross-section of collisional interactions between electrons and ions increases with temperature, the strength of CELs scales with the electron temperature.}

\subsubsection{The intricacies of photoionization models}

Modeling the nebular emission in galaxy SEDs requires detailed photoionization codes, such as CLOUDY \citep{1998PASP..110..761F, 2023RMxAA..59..327C} or MAPPINGS \citep{1996ApJS..102..161D, 2022ApJ...927...37J}, which take into account the complex physics of ionized gases, including radiation transfer, heating and cooling processes, and chemical networks. 
Photoionization models typically assume a Str\"{o}mgren sphere \citep{1939ApJ....89..526S}, a fully ionized sphere of uniform density with a central ionizing source. The inner radius of the sphere is set by ionization balance between the rate of photoionization and recombination, and known as the Str\"{o}mgren radius. This is itself dependent on the ionizing source, and the hydrogen density \citep{2001MNRAS.323..887C, 2016MNRAS.462.1757G, 2016MNRAS.456.3354F}.

Photoionization models introduce a number of additional free parameters, including the dimensionless ionization parameter, the hydrogen density, the Lyman-continuum escape fraction, and those describing the geometry of the cloud, as well as the abundance patterns of the ionizing source and the surrounding cloud, and the depletion patterns of those elements onto grains of different sizes. Many modern SPS codes self-consistently include nebular emission \citep{2017ApJ...840...44B}. This is done in two steps, (i) computing the number of ionizing photons from the star formation history with an assumed stellar population model, followed by (ii) using precomputed tables from photoionization codes to generate a corresponding nebular emission spectrum that is used to produce a composite stellar plus nebular spectrum.

\subsubsection{Radio emission from plasma in the ISM}

Radio emission from plasma in the ISM has two primary components, both of which trace star formation \citep[see e.g.,][for reviews]{condon92, padovani2016}. The dominant component at frequencies below $\sim 30$ GHz is synchrotron emission, produced by cosmic ray electrons (accelerated in supernova remnants) spiraling in galactic magnetic fields. They often show a characteristic power-spectrum of $S_\nu \propto \nu^\alpha$ with $\alpha \approx -0.7$. At higher frequencies ($\gtrsim 30$ GHz), thermal free-free emission becomes significant. This Bremsstrahlung radiation comes from electron-ion collisions in HII regions around massive stars, and is characterized by a nearly flat spectrum ($\alpha \approx -0.1$). Both can be used to probe star formation by probing the rate of core-collapse supernova or the rate at which ionizing photons are being produced by massive stars, and have the advantage of being relatively unaffected by dust, which can be useful for probing the fraction of star formation obscured by dust at a given epoch. 
\ch{With many current and upcoming wide-field, deep surveys using sensitive radio interferometers (LOFAR, Meerkat, NgVLA, SKAO), the radio frequency regime is a promising window to efficiently achieve a dust-unbiased census of star formation across cosmic time}.

\subsection{Dust in the Interstellar Medium}
\label{sec:dust}

\begin{figure}
    \centering
    \includegraphics[width=\linewidth, page=5, trim={0 21cm 0 0},clip]{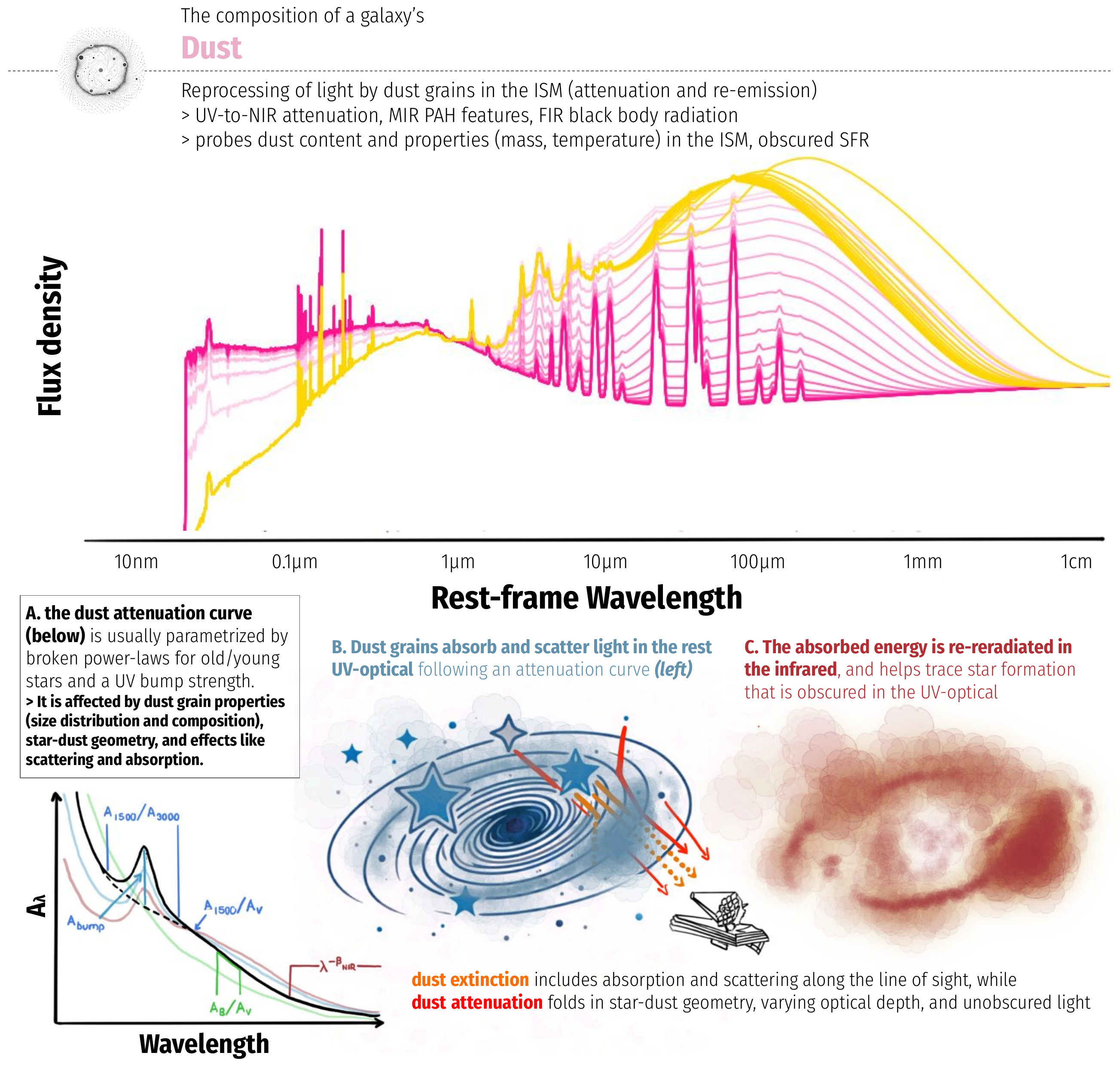}
    \caption{Stellar and nebular light is reprocessed by dust. \textbf{Top:} A galaxy spectrum with varying levels of attenuation (pink) and with varying parameters for the dust re-emission (yellow). \textbf{Bottom:} the dust attenuation curve parameters (A) based on the review by \cite{2020ARA&A..58..529S}, and a schematic showing dust attenuation in the optical-UV and re-emission in the infrared.}
    \label{fig:sed_anatomy_dust}
\end{figure}

Cosmic dust particles (or grains) form in the atmospheres of evolved stars and are released into the ISM by stellar winds or explosions \citep{Draine2011}. Dust grains typically range in size from 5 to 250 nm \citep{Weingartner2001}, which is much smaller than what we call dust on Earth. Dust affects the SED of galaxies primarily in two ways: by attenuating the light at UV-to-NIR wavelengths and by re-emitting light in the IR following energy balance. Here we take a look at the SED features of dust attenuation and re-emission and we refer the reader to reviews to get a more in-depth picture (e.g., \citealt{10.48550/arXiv.astro-ph/0008403, 10.48550/arXiv.2211.00850, 2020ARA&A..58..529S}). SEDs of galaxies affected at varying levels of dust attenuation and varying re-emission temperatures are represented in Figure~\ref{fig:sed_anatomy_dust}.

\subsubsection{Attenuation at UV, optical, and NIR wavelengths}

Dust affects the SED of galaxies by ``absorbing'' the light at UV, optical, and NIR wavelengths. To first order, the effect depends on the size of the dust grains. Grains of small size tend to be more abundant than grains of large size, thus shorter wavelengths are more affected than longer wavelengths, making the SED ``red''.

``Extinction'' and ``attenuation'' are common terms when describing the effect of dust on the SEDs of galaxies. Although they are sometimes used interchangeably, there is a difference between the two mechanisms. Extinction represents the amount of light absorbed or scattered away along a single line of sight through a dusty medium. Attenuation is a combination of extinction plus scattering of light back into the line of sight and the additional contribution by unobscured stars. A schematic representation is shown in Figure \ref{fig:sed_anatomy_dust}. The attenuation depends on how the light interacts with the dust as it makes its way out of the galaxy and toward our telescopes. Factors such as the amount of dust surrounding the birth-clouds of young stars, \ch{grain-size distributions, the ratio of silicates and carbonate grains,} the clumpiness of dust in the ISM, and even galaxy-wide factors like the inclination of galaxies with respect to the observer can affect the amount of differential attenuation affecting the stellar populations in any given galaxy. Modern simulations use radiative transfer methods like SKIRT \citep{2003MNRAS.343.1081B, 2015A&C.....9...20C} and POWDERDAY \citep{2021ApJS..252...12N} to generate realistic galaxy SEDs by forward modeling the effects of starlight interacting with dust for different attenuation laws and geometries. \ch{These dust radiative transfer approaches are also used to model the effect of dust in idealised situations in the local Universe, as part of inverse modelling processes.}

\textbf{Other sources:} In addition to dust associated with star forming molecular clouds and interstellar clouds, dust can also be produced by material ejected from evolved stars, such as asymptotic giant branch (AGB) stars, with the dust composition being carbonaceous and silicate depending on whether the stars are carbon- or oxygen-rich respectively \citep[see review by][]{2020A&A...641A.103V}. 

\subsubsection{Emission at mid IR, far IR, and sub mm wavelengths}

Dust is heated by short wavelength photons and re-emits light at long wavelengths. About half of all the UV/optical light produced by stars in the Universe is reprocessed by dust \citep{Dole2006}. Dust emission can be represented as a combination of out-of-equilibrium polycyclic aromatic hydrocarbons (PAHs) that emit in the mid-IR \citep{Draine2001} and black body emissions at different temperatures in the mid- to far-IR (see e.g., the models by \citealt{Chary2001, 2002ApJ...576..159D, 2008MNRAS.388.1595D}).
The emission in the mid-IR (continuum and emission lines) comes from polycyclic aromatic hydrocarbons (PAHs) heated in the photodissociation regions around young stars. The most prominent features are at wavelengths 3.3 $\mu m$, 6.2 $\mu m$, 7.7 $\mu m$, 8.6 $\mu m$, and 11.3 $\mu m$ and are caused by $C-C$ and $C-H$ stretching and bending modes \citep{2019ApJ...884..136X, 2019arXiv190310397S, 2022MNRAS.510.4888K, 2024A&A...685A..75C}. 
The far-IR emission is characterized by a bump that depends on the temperature of warm and cold grains in thermal equilibrium. The most prominent emission lines are from fine-structure transitions of atoms like carbon ([C II] at $\sim$160 $\mu m$) and oxygen ([O III] line at $\sim$90 $\mu m$) \citep{Stacey1991, Brauher2008}.

Dust emission observations are used to constrain the dust mass and temperature in galaxies and also the star formation rate. This is because the light from young stars is attenuated and re-emitted by the dust and dust-obscured star formation contributes significantly to the total star formation rate density of the Universe (e.g., \citealt{2014ARA&A..52..415M}). The connection between dust emission and star formation rate has been empirically calibrated for simple conversions. The most notable emission lines in the FIR spectrum arise from fine-structure transitions of atoms. The [C II] fine structure line at approximately 158$\mu$m is one of the brightest and most studied lines in this region, and is considered a dominant coolant of the ISM gas \citep{2017ApJ...846...32D, 2023A&A...679A.131R, 2024ApJ...965..179G}.

\subsubsection{\ch{Anomalous Microwave Emission (AME)}}

\ch{Emission from galaxies detected in radio frequencies (typically $\sim 10-60$ GHz) is thought to arise from the electric dipole radiation from rapidly spinning small dust grains \citep[see the review by][]{2018NewAR..80....1D}. Though it correlates with far-IR thermal dust emission, it was considered anomalous since it could not be fully explained by conventional mechanisms such as synchrotron or free-free radiation.}

\subsection{Active Galactic Nuclei}\label{sec:agn}

\begin{figure}
    \centering
    \includegraphics[width=\linewidth, page=6, trim={0 33cm 0 0},clip]{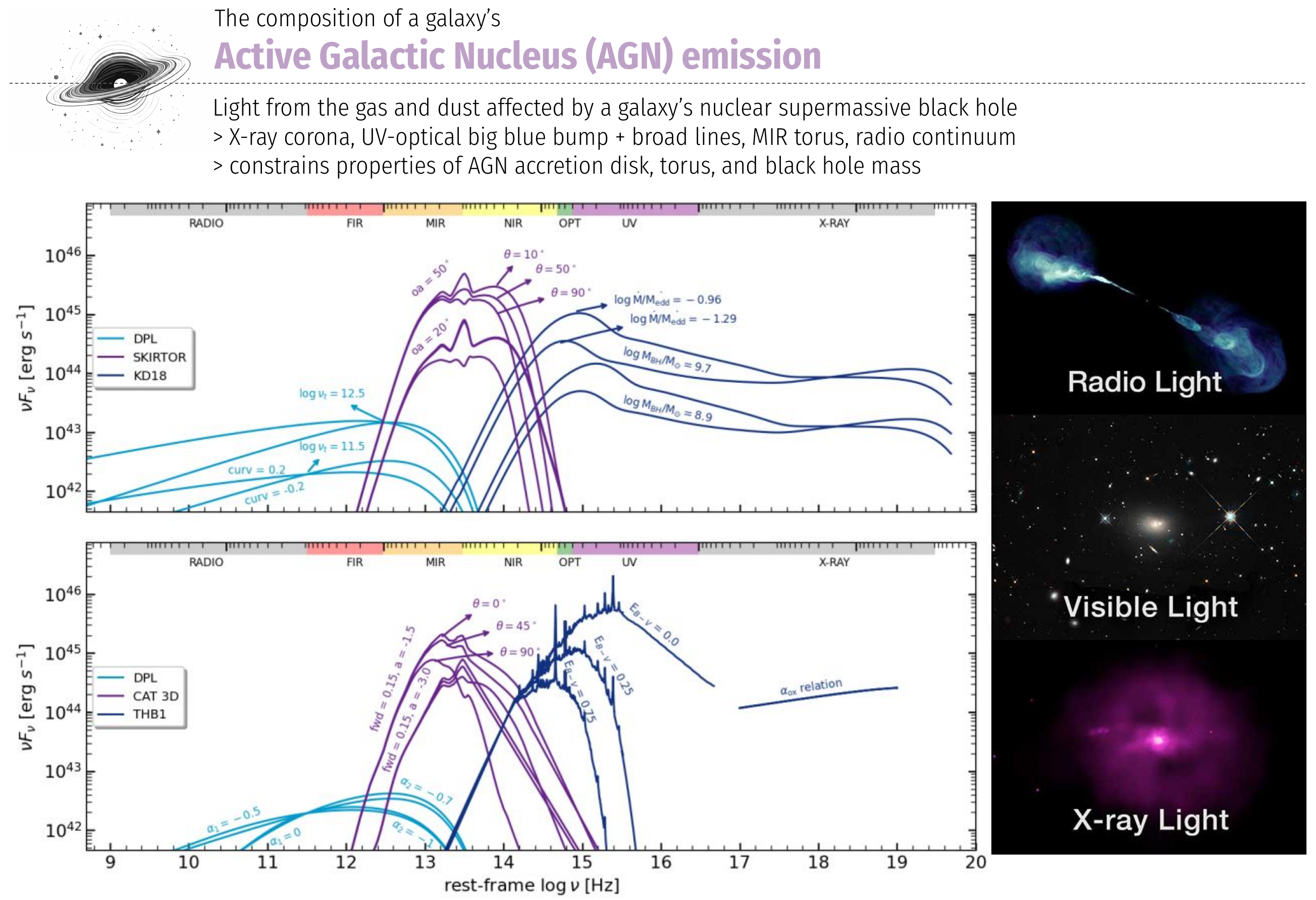}
    \caption{\textbf{(Left:)} Models of the AGN emission across the electromagnetic spectrum. Theoretical and empirical SED models for the accretion disk and X-ray emission (dark blue), the hot nuclear dust torus emission (purple), and radio synchrotron (light blue) are shown for comparison, adapted from \cite{2024A&A...688A..46M}.  \textbf{(Right:)} An image of Hercules A at different wavelengths illustrates AGN emission (X-ray: NASA/CXC/SAO; visual: NASA/STScI; radio: NSF/NRAO/VLA).}
    \label{fig:sed_anatomy_agn}
\end{figure}

There is evidence that most, if not all, massive galaxies host a massive or supermassive black hole (SMBH) in their center. When the SMBHs are active, i.e. actively growing through the accretion of matter, they are known as active galactic nuclei (AGN) and can be very luminous across the entire electromagnetic spectrum. Depending on the AGN power, the presence of an AGN can significantly alter the overall observed SED of a galaxy, or even completely dominate the emission across the spectrum in the most extreme sources such as quasars and radio galaxies \citep[see reviews by][]{1993ARA&A..31..473A, 2000ARA&A..38..521S, 2012ARA&A..50..455F, 2013ARA&A..51..511K, 2014ARA&A..52..529Y, 2015ARA&A..53..365N}.
Here we summarize the main physical components of the nucleus of an AGN, in order of increasing wavelength regime in which they dominate. Models of these components are also shown in Figure \ref{fig:sed_anatomy_agn}.

\subsubsection{High-energy black hole physics: optical, UV, and X-rays}

At distances ranging from a few light-days to several light-years around the central black hole, surrounding gas and matter fall onto the gravitational potential of the black hole forming an optically thick accretion disk. This accretion disk emits luminous thermal radiation that peaks in the ultraviolet and optical regions of the spectrum, often referred to as the ``big blue bump'' \citep{1989ApJ...347...29S}.
The observed SED of the accretion disk is further influenced by factors such as dust attenuation at both nuclear and galactic scales, as well as the presence of broad (hundreds of km/s) and narrow (300–1,000 km/s) emission lines. These emission lines originate from ionized gas clouds located at distances of 1 to several hundred parsecs from the accretion disk \citep[see e.g.,][]{2011Natur.470..366K}.
The wavelength regime of the accretion disk emission coincides with the stellar continuum and is therefore most easily identifiable in luminous active galactic nuclei (AGN) that are relatively free of dust attenuation, such as Type-1 AGN or quasars, where the accretion disk can outshine emission from the host galaxy by orders of magnitude \citep{2018MNRAS.480.1247K}.

Above the accretion disk, a corona of hot electrons generates X-ray continuum emission with a power-law shape via inverse Compton scattering of ultraviolet photons emitted by the accretion disk. This X-ray emission is further modified by various nuclear components, including photoelectric absorption, reflection, and scattering \citep{1993ApJ...413..507H, 2017MNRAS.467.2566F}.
Although star formation can also produce X-ray emission, it is generally much weaker than the X-ray output from an AGN. As a result, the X-ray regime serves as one of the most reliable tracers for identifying AGN activity.

\subsubsection{Hot nuclear dust: infrared}

A significant fraction of the total AGN energy is emitted in the NIR and MIR regimes. This emission originates from nuclear material composed of hot gas and dust, located at scales of a few parsecs, which is heated by the radiation from the accretion disk. Commonly referred to as the AGN torus, this structure's precise geometry and composition remains a topic of ongoing debate \citep{2012MNRAS.426..120F, 2019ApJ...884...10G, 2024arXiv241010941T}. Recent studies suggest that it consists of a clumpy toroidal configuration, potentially with additional polar components.
The gas within this structure serves as the primary source of X-ray obscuration, while the dust attenuates the emission from the accretion disk. The morphology and orientation of this dusty component play a crucial role in shaping the observed diversity of AGN. For instance, the classification of AGN as Type 1 or Type 2 is strongly linked to the viewing angle of this toroidal structure: a face-on view corresponds to Type 1 AGN, while an edge-on view corresponds to Type 2 AGN.

\subsubsection{Interacting plasma: radio}
AGN emission in the radio regime is primarily driven by synchrotron processes, which involve the interaction of a plasma of relativistic electrons with a magnetic field. These processes can originate from various phenomena, including jets, AGN wind shocks, and the X-ray corona. Radio jets, in particular, are relativistic, collimated streams of plasma launched by the kinetic energy released during the accretion process. The sizes of AGN jets vary widely, ranging from sub-galactic scales to cosmological extents of several megaparsecs (Mpc). However, large-scale jets are observed in only a minority of AGN, approximately 10\%. When the jet is aligned close to the line of sight (blazars) the synchrotron SED from the jet can dominate not only the radio regime but also the entire electromagnetic spectrum.

\subsection{The Circumgalactic Medium (CGM)}

The circumgalactic medium (CGM) consists of multiphase gas extending out to the virial radius of galaxies \citep[see reviews by][]{2017ARA&A..55..389T, 2023ARA&A..61..131F}, and can influence galaxy SEDs by scattering and absorbing light at UV-optical wavelengths and through emission from collisionally excited gas. Gas in the CGM is a mixture of pristine gas from inflows and enriched gas from feedback-driven outflows \citep{2020ApJ...903...32F}, and is a key tracer of baryon cycling in galaxies \citep{2017MNRAS.470.4698A}. The extended CGM around galaxies can be studied by examining absorption lines (MgII, C IV, and O VI) in the SEDs of bright background sources \citep{2013MNRAS.432...89F, 2015ApJ...802...10C, 2023MNRAS.519.5514A}, or by studying the faint emission (Ly-$\alpha$, [OII], and [OIII]; \citealt{2019MNRAS.489.2417A, 2020ARA&A..58..617O, 2021MNRAS.501.5757M}) using telescopes like Dragonfly \citep{2014PASP..126...55A}, FIREBall-2 \citep{2022arXiv221115491P} and HARMONI \citep{2014SPIE.9147E..25T}.

\subsection{The Intergalactic Medium (IGM)}

The intergalactic medium (IGM), composed primarily of clouds of hydrogen and helium, can also affect the SEDs of galaxies, particularly at high redshifts, through absorption and scattering of light \citep[see reviews by][]{1998ARA&A..36..267R, 2005ARA&A..43..861W, 2016ARA&A..54..313M, 2023ARA&A..61..373F}. These can be primarily broken down into three classes: Lyman-alpha forest (LAF), Lyman-limit systems (LLS) and Damped-Lyman Absorbers (DLAs). Though studied primarily in the context of producing absorption features along lines of sight to distant bright objects like quasars\footnote{which allows us to trace the cosmic structure along these lines of sight.}, they are also important in accurately modeling the far-UV spectra of galaxies and correctly estimating distances using the Lyman-break technique. DLAs are also particularly interesting in the context of high-z JWST observations, where extended gas around galaxies can give rise to smooth damping wings that soften the Lyman break\footnote{which could be used to trace the structure of reionization bubbles \citep{2024arXiv241104176L}.} \citep{1998ApJ...501...15M, 2024arXiv241021543A, 2024ApJ...971..124U, 2024Sci...384..890H}. 

\section{\ch{Physical properties inferred from galaxy SEDs}}
\label{sec:sedprop}

In the previous section we described the information about a galaxy's history that is encoded in its SED, which can be gleaned from detailed modeling of their stellar, nebular, dust and AGN components. However, since these different components often overlap in the wavelength ranges of their emission, we need to be careful in how we approach the inverse problem of determining galaxy properties from their spectra. In this section, we describe the physical properties that are traditionally determined from a galaxy's SED, and in the following section, we describe the techniques used to constrain these properties using modern multi-wavelength observations.

\subsection{Redshifts and distance measurement}
\label{sec:redshift}

The measurement of cosmic distances through redshifts underpins most of modern observational galaxy evolution and cosmology. As light travels across expanding space, its wavelength stretches in proportion to the scale factor of the universe, providing a key observational signature of cosmic expansion and a means to probe vast astronomical distances \citep{1929PNAS...15..168H, 1987ApJ...313...42D}.

\subsubsection{Spectroscopic redshifts (spec-z)} 

Spectroscopic redshift measurements represent the gold standard in distance determination, offering precise measurements through the identification of known spectral features such as strong emission lines and Lyman- or Balmer-breaks. By measuring the wavelength shifts of emission and absorption lines relative to their rest-frame values, spectroscopic observations can typically achieve precision of $\Delta z / (1+z) \sim 10^{-4}$ for high SNR spectra.

\textbf{grism-z:} Slitless spectroscopy, or \textit{grism}, while offering the advantage of simultaneous spectral measurements for many objects, presents unique challenges for redshift determination. The low spectral resolution and wavelength-dependent contamination from nearby sources can lead to degeneracies in line identification. Furthermore, the slitless nature of grism observations means that extended sources produce spectra convolved with their spatial profiles. \ch{While this requires the construction of specialized extraction and analysis techniques \citep[e.g., aXe, pyLinear and grizli][]{2009PASP..121...59K, 2018PASP..130c4501R, 2021zndo...5012699B}, which have been used across a range of different science cases, ranging from spatially resolved emission line maps \citep{2018ApJ...868...61P} to redshifts and star formation history inference \citep{2022MNRAS.512.3566N, 2024MNRAS.532..577E, 2024A&A...690A..64M}.}

\subsubsection{Photometric redshifts (photo-z)}

While spectroscopic observations are the preferred method for determining accurate distances to galaxies, they are expensive due to the steep exposure times required as we consider fainter or more distant objects. Photometric redshift techniques provide a valuable alternative. These methods leverage broad-band photometry to estimate redshifts, trading precision for observational efficiency and reaching much larger source populations. As precise photometric redshifts are increasingly needed for the next generation of cosmological experiments, there have been rapid advances in techniques to infer distances using both traditional template-based methods \citep[EAZY][]{2008ApJ...686.1503B, 2023zndo...8268031B} and newer methods based on machine learning and astrostatistics. \cite{2013ApJ...775...93D, 2020MNRAS.499.1587S, 2020A&A...644A..31E, 2023ApJ...942...36K} compare different photo-z methods and their effect on estimated redshifts in galaxy surveys. The review by \cite{2022ARA&A..60..363N} provides a in-depth summary of current techniques, described briefly below.

\textbf{Template based methods:} Template fitting approaches compare observed photometry to a library of spectral templates evolved through redshift space. The likelihood of each template-redshift combination is evaluated, typically incorporating prior information about galaxy populations such as luminosity or apparent magnitude in a given band. While conceptually straightforward, these methods are sensitive to template completeness and calibration, particularly at high redshifts where templates may be less representative. Some commonly applied methods that use a $\chi^2$-based likelihood include LePhare, BPZ, and EAZY. 

\textbf{Data driven or empirical methods:} Template-based methods may fail in cases where the template set is incomplete, the prior is unknown or incorrect, or the data do not have enough SNR to constrain the model. In these cases, statistical and machine learning techniques have emerged as powerful tools for photometric redshift estimation, sacrificing some interpretability while learning complex mappings between galaxy SEDs and redshifts. Methods ranging from random forests \citep{2013MNRAS.432.1483C}, Gaussian processes \citep{2016MNRAS.455.2387A} and manifold-based methods \citep{2014MNRAS.438.3409C, 2015ApJ...813...53M, 2017MNRAS.469.1186S} to neural networks \citep{2003MNRAS.339.1195F, 2021AJ....162..297L} can learn these mappings between photometric features and redshifts when trained on labeled or semi-labeled samples (the labels here often being spectroscopic redshifts or an equivalent gold standard), and also incorporate additional data such as information about galaxy morphologies \citep{2023MNRAS.518.5049H, 2024A&A...683A..26A}. These approaches can achieve superior precision compared to template fitting, but their reliability depends critically on the representativeness of training sets across the color-magnitude space of interest. Some ML-based photo-z methods include ANNz \citep{2004PASP..116..345C}, Delight \citep{2017ApJ...838....5L}, FlexzBoost \citep{2017arXiv170408095I}, METAPHOR \citep{2017MNRAS.465.1959C} and DNF \citep{2024A&A...686A..38T} and hybrid template+ML methods like HAYATE \citep{2024MNRAS.530.2012T}.

\subsection{Stellar Masses}

Stellar masses are perhaps one of the most robust quantities that can be derived from SED fitting, with the least wavelength coverage / SNR. The total stellar mass $M_*$ is determined by fitting observed SEDs with the understanding that the integrated (stellar and nebular) light of the galaxy is composed of contributions from stellar populations of different ages (i.e. SSPs) weighted by the star formation rate (SFR) at that age, which can be written as 
\begin{equation}
F_{obs}^i = \frac{1}{4\pi D_L^2} \int_0^{t_{obs}} SFH(t') F_\lambda (t-t', Z) e^{-\tau_\lambda(t')} dt'
\end{equation}
where SFH(t) is the star formation history, $F_\lambda$ is the spectrum of an SSP at a given age and metallicity, $\tau_\lambda$ is the dust attenuation, and $D_L$ is the luminosity distance
This makes several assumptions, such as the form of the IMF, and the treatment of dust and metallicity, but generally provides robust masses as long as the SED can probe the portion of the rest-optical that is sensitive to older stars (usually between $4000$\AA{} and $1.6\mu$m). Typical uncertainties range from $0.2-0.3$ dex, dominated by systematic effects rather than photometric errors.

A major systematic in measuring the stellar mass comes from the increased sensitivity of a galaxy SED to light from young, massive stars, an effect sometimes called outshining, which can cause traditional fitting methods to underestimate the contribution from older stellar populations and therefore the total stellar mass by factors of $\sim 0.1-0.3$ dex\footnote{It is important to note that this statement is made in the context of UV-to-NIR SEDs, and can actually lead to overestimated stellar masses with traditional priors for high-z observations that do not have rest-NIR coverage \citep{2023ApJ...958...82S}.} \citep{2009ApJS..184..100L, 2015MNRAS.452..235S, 2015MNRAS.446.1512H, 2017ApJ...838..127I, 2018ApJ...853..131L, 2019ApJ...873...44C, 2019ApJ...876....3L, 2020ApJ...904...33L, 2023ApJ...944..141P, 2024arXiv240903959D, 2024MNRAS.527.3291J}.  This can be mitigated in several ways, most notably by adopting appropriate priors with non-parametric SFHs \citep{2020ApJ...904...33L}, or by studying resolved observations \citep{2013ApJ...770...57B, 2015MNRAS.452..235S, 2015MNRAS.446.1512H, 2017ApJ...838..127I, 2018MNRAS.476.1532S, 2023ApJ...958...82S, 2024A&A...686A..63G}, where the fact that young stars tend to be clustered allows us to better capture the older populations, or by accurately capturing the stochasticity of star formation rates \citep{2023MNRAS.526.2665S, 2024MNRAS.532.4002W, 2024ApJ...961...73N, 2024MNRAS.530L...7H, 2024ApJ...975..192G, 2024arXiv240707937W}. 

Robust constraints on stellar masses propagate to all levels of studies of galaxy evolution, since ensemble statistics including stellar mass functions at different redshifts and scaling relations like the SFR-M$_*$ correlation are often used to test and calibrate cosmological simulations and other models of galaxy evolution. Stellar mass functions and stellar-halo mass relations (SMHM) are key to understanding the average growth of galaxies across redshifts \citep{2015ARA&A..53...51S, 2016ApJ...833....2G, 2017MNRAS.470..651R, 2019MNRAS.488.3143B}, while scaling relations like the SFR-M$_*$ correlation and the mass-metallicity relation may help us better understand how galaxies self-regulate star formation, undergo chemical enrichment, and eventually quench \citep{2004MNRAS.351.1151B, 2011MNRAS.412.1123P, 2012ApJ...745..149L, 2014arXiv1406.5191K, 2015ApJ...799..183S, 2016ApJ...832....7A, 2017MNRAS.472.2054P, 2019MNRAS.490.3234N, 2019ApJ...872..160H, 2020ARA&A..58..157T, 2021ApJ...907..114D, 2021ApJ...910...87K, 2024arXiv240613554R}. 

Many uncertainties still remain in the accurate inference of stellar masses, ranging from data-level issues like photometry in crowded fields, to systematics in SPS templates such as the IMF and uncertainties in current models of stellar atmospheres, to the burstiness of galaxy-wide SFHs, especially at high redshifts. As we look forward to the next generation of galaxy surveys, these uncertainties will be better quantified and accounted for with a combination of theoretical advances, targeted observations and data-driven modeling.

\subsection{Star formation rates}

The star formation rate (SFR) of a galaxy is one of its most fundamental properties, tracing the rate at which gas is able to cool and form stars at the time of observation.
It is accessible from several parts of its SED, ranging from stellar and nebular emission from young, short-lived stars in the UV and optical to re-emission from dust in the infrared. For a detailed treatment, see reviews by \citealt{2012ARA&A..50..531K, 2014ARA&A..52..415M}. A key advantage of full SED fitting is the ability to simultaneously constrain both unobscured and dust-obscured star formation using observations across the electromagnetic spectrum. However, many studies still prefer to convert the strength of specific spectral features into a SFR for its inherent simplicity and ease of interpretation, the most common of which are described below. While we have reported established calibrations from \cite{Kennicutt1998} here unless otherwise specified, updated calibrations such as \cite{2007ApJ...666..870C, 2011ApJ...737...67M, 2022ApJ...929....3C, 2024ApJ...970...61R} are being developed with modern surveys that incorporate more careful treatments of different physical mechanisms and higher SNR observations.

\begin{enumerate}
    \item \textbf{Nebular emission} lines, particularly from H$\alpha$ and other Balmer lines directly traces ionizing radiation from massive stars, and probes star formation on short timescales $\sim 4-10$ Myr \citep{2021MNRAS.501.4812F} after correcting for dust using the Balmer decrement. SFR(M$_\odot$/yr$)_{H\alpha} = 7.9(10^{-42}) $L$_{H\alpha}$ (erg/s) or SFR(M$_\odot$/yr$)_{[OII]} \approx 1.4(10^{-41}) $L$_{[OII]}$ (erg/s).
    \item \textbf{UV continuum} ($1500-2800$\AA) is sensitive to emission from short-lived massive stars $M_\odot \gtrsim 5 M_\odot$, and also probes short-timescale SFR ($\sim 30-70$ Myr) \citep{2021MNRAS.501.4812F, 2023A&A...673A..16K}. The canonical calibration relates UV luminosity to SFR assuming continuous star formation and a Salpeter IMF as SFR(M$_\odot$/yr$)_{UV} = 1.4(10^{-28}) $L$_\nu$ (erg/s/Hz). Dust in the UV can significantly complicate this, and requires careful correction before converting the luminosity to SFR. 
    \item The total \textbf{IR luminosity} $\sim 8-1000\mu$m provides a robust SFR indicator by tracing dust re-emission, since it captures the bulk of the energy from young stars and thus traces $\sim 100$ Myr timescales. Similar to the UV, there exist calibrations to convert the IR luminosity to SFR, given by SFR(M$_\odot$/yr$)_{IR} = 4.5(10^{-44}) $L$_{IR}$ (erg/s). 
    However, this assumes that the IR emission comes solely from young stellar populations, and thus might overestimate SFR in galaxies with significant heating of dust from older stars\footnote{updated calibrations exist that take this into account such as \citep{2011ApJ...737...67M} SFR(M$_\odot$/yr$)_{24\mu m} = 5.58(10^{-36}) \nu $L$_{\nu, 24\mu m}^{0.826}$ (erg/s)}. Mid-IR PAH features (e.g. at 7.7 $\mu$m), fine-structure lines in the IR such as [Ne II] $12.8\mu$m and [Ne III] $15.6\mu$m, and [CII] $158\mu$m emission associated with cooling in photodissociation regions can also be used to independently trace the SFR.
    \item \textbf{Rest-optical Balmer absorption} features like H$\delta$ absorption and the $D4000$ break strength trace SFR out to longer timescales \citep[$\sim 100$ Myr to $1$Gyr; ][]{1997A&A...325.1025P, 2004MNRAS.351.1151B, 2020A&A...633A..70P}. 
    \item \textbf{Radio continuum} emission (e.g. at $1.4$ to $33$GHz) from supernova remnants and cosmic rays is not affected by dust and can trace SFRs over intermediate to long timescales depending on the precise frequency being measured and its underlying physical mechanism \cite[e.g. $\sim 200-300$ Myr at 1.4GHz; ][]{2011ApJ...737...67M, 2023A&A...675A.126A, 2024MNRAS.531..708C}.
\end{enumerate}

For galaxies with significant obscuration by dust, the rest-UV or nebular emission based indicators might only provide a lower estimate of the SFR, even when corrected for dust. This is especially true when the dust is clumpy, and light is only able to leak out of regions with lower optical depth. In these cases, IR-based estimators or emission lines that are relatively unaffected by dust like Pa-$\alpha$ are key indicators of the total SFR, and help determine how much of the SFR is obscured by dust. At high redshifts, this fraction of dust-obscured SFR remains an open question.

Additionally, the assumed choice of star formation history impacts the inferred SFR, especially for indicators that probe SFR on longer timescales. Generally, assuming more flexible SFHs leads to higher stellar masses and lower SFRs from priors on the amount of light attributed to older stellar populations. Modern approaches often use SPS models to derive calibrations specific to the galaxy population being studied, rather than applying universal conversion factors. This is particularly important at high redshift where conditions may differ significantly from local galaxies \citep{2021MNRAS.501.4812F, 2024ApJ...961...53I}.

\subsection{Star Formation Histories}

Decoding the imprints of past star formation activity in the rest-UV to NIR SEDs unlocks access to a whole host of studies tracing the evolution of individual galaxies through time. Since the SFHs of galaxies contain the imprints of the many physical processes that regulate star formation in galaxies, this provides some of the most sensitive constraints on baryon cycling and feedback possible at the population level. Historically, this has been a hard problem due to (i) emission from non-stellar sources that need to be modeled and disentangled, (ii) the rapid decrease in sensitivity for stellar populations older than $\sim 1$Gyr, (iii) the lack of large pan-chromatic datasets with enough SNR to disentangle subtle differences in stellar populations, (iv) flexible descriptions that provide an information-rich representation of galaxy SFHs, (v) a better understanding of systematics due to SPS, dust and AGN models, and (vi) the computational difficulties in sampling the high-dimensional parameter space required to simultaneously constrain all the free parameters from a flexible SFH. The last decade has seen several improvements in all these areas, in particular with the rise of public SED fitting codes and implementations and priors for non-parametric star formation histories, which have enabled the reconstruction of detailed information not only of the galaxy's mass- and light-weighted ages, but also of the overall shape of galaxy SFHs.

\textbf{Analytic SFHs:} Analytic SFHs have the advantage of being simple and computationally efficient, but do not capture the full complexity of real SFHs. Despite possible biases, a significant fraction of the literature still uses parametric forms to describe galaxy SFHs \citep{2019ApJ...873...44C}, ranging from constant SFHs and exponentially declining ($\tau$-model) SFHs, to slightly more complex SFHs such as those with linear or polynomial rise followed by exponential decline, lognormal, or double-power law SFHs, to more physically motivated SFHs either based on evolution along the star forming main sequence \citep{2017A&A...608A..41C} or SFH parametrizations based on gas depletion and star formation efficiency \citep{2023MNRAS.518..562A}. 

\textbf{Non-parametric SFHs:} Non-parametric SFHs provide more flexibility in describing complex star formation histories and their priors. However, they require more free parameters and can be more computationally expensive to fit. Several non-parametric descriptions of galaxy SFHs have been used to describe the overall shape of the SFH, including SFRs binned in time \citep{2000MNRAS.317..965H, 2007MNRAS.381.1252T, 2017ApJ...837..170L}, or in quantiles of mass formed \citep{2019ApJ...879..116I}, SFHs based on semi-analytic models \citep{2013ApJ...762L..15P}, and SFHs described as a sum of Gaussians \citep{2020MNRAS.495..905R} or polynomials \citep{2022A&A...662A...1J}. In addition to the specific form of the SFH adopted, non-parametric SFHs also allow for custom prior distributions that can be tailored to include known information about the galaxy populations being modeled. \ch{A major impact of the adoption of non-parametric SFHs has been the reduction of biases in estimating SFH-related quantities like stellar mass, SFR and formation times due to the imposition of a functional form and its priors \citep{2015MNRAS.446.1512H, 2015MNRAS.452..235S}. This generally leads to more massive stellar masses, older ages, and slightly lower SFRs compared to parametric approaches \citep{2017ApJ...838..127I, 2020ApJ...904...33L, 2019ApJ...877..140L}.} In recent times, non-parametric SFHs have also been used in conjunction with machine learning based methods to improve robustness and speed up inference \citep{2019MNRAS.490.5503L, 2023ApJ...954..132M, 2024A&A...689A..58I, 2024arXiv240807749G}.

\subsubsection{Burstiness} 

Star formation in nearby galaxies and in simulations tends to be quite stochastic. This stochasticity is driven by a range of mechanisms including galaxy interactions, gas accretion events, and baryon cycling. Accurately modeling and constraining the `bursty' behavior of star formation is important for robustly inferring properties from the SEDs of some galaxies (particularly at high redshifts), and can be done by comparing the distributions of SFR indicators sensitive to star formation on different timescales \citep{2021MNRAS.501.4812F, 2024ApJ...961...53I, 2024ApJ...975..192G, 2023MNRAS.526.2665S, 2024MNRAS.532.4002W}. The studies often use measurements of spectral indicators or star formation rates for a population of galaxies, since SFR fluctuations are not accessible through individual galaxy SFHs. 
Modern IFU surveys and other spatially resolved studies of star clusters have revealed that this burstiness often has spatial structure within galaxies \citep{2018MNRAS.480.2544R, 2020ApJ...892...87W, 2023ApJ...952..133M, 2024MNRAS.532..577E}. 

\subsubsection{Spatially Resolved Stellar Populations}

The advent of wide-field IFU spectroscopy from surveys like SDSS-IV MaNGA \citep{2015ApJ...798....7B}, SAMI \citep{2015MNRAS.447.2857B, 2021MNRAS.505..991C} and CALIFA \citep{2012A&A...538A...8S, 2013A&A...549A..87H} (and high-resolution imaging surveys at higher redshifts) have enabled detailed mapping of stellar populations within galaxies. Spatially resolved studies reveal how stellar ages, metallicities, star formation histories, and dust content vary across galactic structures like bulges, disks, and spiral arms. This spatial information provides insights into the physical processes that drive galaxy assembly and quenching. For instance, early-type galaxies typically show negative metallicity gradients (the bulges are more metal-rich) but relatively flat age gradients, suggesting they assembled inside-out. In contrast, star-forming disk galaxies often display negative age gradients (younger outer regions) alongside negative metallicity gradients, consistent with continuing inside-out growth through gas accretion. Studies of transitioning ``green valley'' galaxies reveal diverse quenching patterns: some show evidence for outside-in quenching with star formation ceasing first in their outskirts, while others display inside-out quenching progressing from their centers. These patterns can distinguish between different physical drivers - for example, inside-out quenching often suggests AGN feedback or morphological quenching, while outside-in patterns point toward environmental processes like ram pressure stripping or strangulation. 

High-resolution observations from HST (and now JWST) extend these studies to high redshifts, revealing how internal galaxy structure develops over cosmic time. These studies show that the correlation between local stellar mass density and stellar age - the ``inside-out growth" pattern - was already in place by $z\sim 2$. However, the diversity of spatially resolved star formation histories increases at high redshift, with some galaxies showing evidence for ``compaction'' events that rapidly build up dense stellar cores. Resolved stellar populations also illuminate the connection between morphological features and assembly history. Bulges typically host older, more metal-rich stellar populations compared to disks, but the detailed age and metallicity distributions can distinguish between classical bulges built through mergers and pseudo-bulges grown through secular processes. Similarly, the stellar population properties of spiral arms, bars, and rings provide insights into their formation timescales and their role in driving radial mixing and secular evolution.

Trends in age and metallicity gradients with galaxy mass, environment, and morphology from spatially resolved surveys provide essential constraints for theoretical models of galaxy formation. For instance, the finding that metallicity gradients generally steepen with galaxy mass but flatten in the highest mass galaxies suggests a transition in the relative importance of in-situ star formation versus merger-driven assembly. Next-generation instruments on thirty-meter class telescopes like TMT/ELT will extend these studies to even higher spatial resolution and redshift, potentially resolving individual star-forming regions in high-redshift galaxies and directly observing the sites of mass assembly during the peak epoch of cosmic star formation.

\subsubsection{Quenching}

The mechanisms responsible for the cessation of star formation in galaxies are still not fully understood, and estimating quenching timescales tends to be one of the primary applications of observationally constraining galaxy SFHs \citep{2018MNRAS.480.4379C, 2021MNRAS.506.5108C}. Immediately following quenching, galaxy SEDs retain significant UV emission from the remaining O- and B-stars, but this declines rapidly over $\sim 100$ Myr as the galaxy is not forming any new massive stars. During this phase, strong Balmer absorption lines develop, particularly H$\delta$, as the SED becomes dominated by A-type stars. This post-starburst or ``E+A'' phase provides a clear marker of recent quenching. Over longer timescales of $\sim1-2$ Gyr, the strength of the 4000\AA{} break increases as the stellar population grows older, while the overall SED becomes progressively redder. These various indicators allow us to time-stamp when quenching occurred.

In addition to the timescale, studying stellar populations in spatially resolved observations allows us to constrain the progression of quenching (i.e. whether it starts from the center of the galaxy or the outskirts, as described in the previous section), and thus the mechanisms responsible \citep{2020MNRAS.499..230B, 2021MNRAS.508..219N, 2023ApJ...945..117A, 2024A&A...686A.124L}. Recent work with JWST has observed galaxy quenching at high redshifts $z \gtrsim 5$, where it is difficult to tell apart galaxies that are experiencing a temporary decline in their SFR from those that have completely ceased forming stars.  The SEDs of these high-z quiescent galaxies often show signs of significant dust content and very high stellar metallicities, suggesting intense periods of star formation and chemical enrichment before quenching.

\subsection{Metallicity and line-based diagnostics}

The chemical enrichment of galaxies through successive generations of stellar evolution leaves distinct imprints across their SEDs. Both stellar and gas-phase metallicities profoundly influence a galaxy's UV-to-NIR SED, albeit in different ways. Higher metallicity stars have stronger absorption features and redder continua due to enhanced line blanketing in stellar atmospheres. The gas phase metallicity primarily affects nebular emission, both through direct line emission from metal species and by altering the thermal balance of the ionized medium as metals provide important cooling channels. In addition to this, metal abundances directly impact dust formation and composition, and therefore indirectly affect the SED across a wide range of wavelengths. In addition to chemical enrichment, the nebular emission can also be used to trace ISM conditions like the electron temperature and ionization state, galaxy kinematics, outflows and emission from AGN. 

Measuring stellar metallicities from SEDs typically relies on either targeted absorption line indices or full spectral fitting approaches. The Lick/IDS system of absorption line indices, particularly combinations of magnesium and iron features like MgII, Fe5270\AA, Fe5335\AA{} have historically been the primary tool for stellar metallicity measurements. However, modern full-spectrum fitting techniques can now leverage the entire optical continuum to simultaneously constrain metallicity alongside other stellar population parameters \citep{2018ApJ...854..139C, 2022MNRAS.513.5446Z}. This is particularly powerful when combined with UV features that probe the metal content of massive stars through both photospheric absorption and stellar wind lines (e.g. Si IV, He II; \citealt{1995ApJS...99..173L, 2001ApJ...550..724L, 2022ApJ...930..105S}). The interpretation of stellar metallicities must also carefully account for degeneracies with age and dust, which can have similar spectral signatures in broad-band photometry.

\ch{Gas-phase metallicity measurements primarily utilize emission line diagnostics. 
The direct method determines abundances by first measuring the electron temperature ($T_e$) using ratios of  temperature-sensitive auroral lines (like [OIII]$\lambda$4363) to nebular lines of the same ion (like [OIII]$\lambda \lambda$4959, 5007). While this provides the most reliable abundances, it requires detecting intrinsically weak auroral lines, which require high SNR and become increasingly difficult at higher metallicities.} 
Strong-line methods provide a more observationally tractable alternative using ratios of bright emission lines like R23 (i.e. the ratio ([OII]$\lambda$3727 + [OIII]$\lambda \lambda$4959,5007)/H$\beta$), N2 ([NII]$\lambda$6583/H$\alpha$), and O3N2 ([OIII]$\lambda$5007/H$\beta$)/([NII]$\lambda$6583/H$\alpha$), which are more readily measurable but provide more uncertain metallicity estimates. 
These diagnostics require careful calibration\footnote{derived either empirically (using samples where oxygen abundances have been measured via the direct method), or theoretically computed using photoionization models.} and can be sensitive to ionization conditions and nitrogen abundance variations \citep{2009MNRAS.398..949P}. 
\ch{An important systematic uncertainty in abundance determinations is the `abundance discrepancy factor' (ADF) - where abundances derived from recombination lines are systematically higher than those from collisionally excited lines, typically by factors of $2-3\times$ in HII regions \citep{2007ApJ...670..457G, 2024arXiv241020819N}.}
While the recent availability of rest-optical SEDs at high-z using JWST has enabled the application of these traditional calibrations to galaxies in the early universe, interpretations need to account for potentially different ISM conditions compared to local calibrations. Auroral line measurements with JWST have unlocked new avenues for determining gas-phase metallicities \citep{2023MNRAS.518..425C, 2023ApJS..269...33N}, and for calibrating strong line measurements.

Spatially resolved studies have revealed significant metallicity gradients within galaxies, suggesting that integrated metallicity measurements may need to account for these variations \citep{2013ApJ...765...48J, 2015MNRAS.448.2030H, 2023MNRAS.520.4301B}. Additionally, evidence for non-solar abundance patterns, particularly $\alpha$-element enhancement in massive galaxies, indicates that assuming scaled-solar abundances may be insufficient \citep{2000ApJ...532..430V, 2015MNRAS.449.1177V}. Some newer stellar population synthesis models now include variable abundance patterns \citep{2024arXiv241021375P}, though this remains an active area of development.

Beyond purely abundance-sensitive diagnostics, emission line ratios are also sensitive to  broader physical conditions of the ISM. The [SII] doublet ratio is sensitive to the electron density. The relative strength of high- to low-ionization species, such as the [OIII]/[OII] ratio, trace the ionization parameter (U). The BPT diagnostic, plotting [OIII]/H$\beta$ against [NII]/H$\alpha$, helps distinguish between star formation and AGN ionization sources by probing the hardness of the radiation field. More complex phenomena also leave distinctive spectral signatures: broad emission features from Wolf-Rayet stars indicate the presence of strong winds produced by very massive stars and are sensitive to the high-mass end of the stellar initial mass function; asymmetric and broad emission line profiles can reveal outflows and their properties; and the complex radiative transfer of Ly$\alpha$ photons through neutral hydrogen produces characteristic line profiles that encode information about the geometry and kinematics of the gas. The interpretation of these various diagnostics requires careful consideration of their interdependencies - for instance, how metallicity affects cooling and thus the temperature structure that underlies many line ratios, or how density variations can impact ionization parameter measurements. Modern photoionization modeling increasingly attempts to simultaneously account for all these effects when interpreting observed emission line spectra \citep{2017ApJ...840...44B, 2024MNRAS.534..523C}.

\subsection{Dust content}

In Section \ref{sec:sedmodel}, we saw that dust affects galaxy SEDs primarily through the absorption and scattering of stellar and nebular light in the UV-to-NIR wavelength range, and re-emitting the energy in the IR \citep[see review by][]{2020ARA&A..58..529S}. 

The wavelength-dependent effect of dust attenuation on the UV-to-NIR SEDs of galaxies is parametrized using a variety of dust laws, both theoretical and empirical. 
In general, these are parametrized by a slope and a normalization usually around the $V$ band, although the wavelength of choice can be arbitrary. 
For example, a single component power-law attenuation law would be given by $A_\lambda = A_V (\lambda/5500$\AA$)^{\beta}$, where $A_V$ is the normalization in the $V$ band and $\beta$ is the slope. 
The slope can be a constant or can vary depending on galaxy properties like morphology or inclination. 
A multi-component attenuation law assigns a different slope and normalization to different components that depend for example on the star formation history of the galaxy (see e.g., \citealt{Cardelli:1989, Calzetti2000, 2000ApJ...539..718C, Charlot2001, Kriek:2013, 2016MNRAS.462.1415C}). 
Some attenuation curves feature a ``dust bump'' in the UV around $\sim 2175$\AA{} due to PAHs and other small grains. 
\cite{Noll:2009, Kriek:2013, 2019A&A...622A.103B} found that steeper attenuation curves for high-redshift galaxies have stronger bumps.
This flexibility is important as studies have shown significant diversity in attenuation curves between galaxies \citep{Salim:2018}. In the absence of IR data, the relationship between UV slope and IR excess (the IRX-$\beta$ relation) can provide statistical constraints, though there is significant scatter in this relation \citep{Meurer:1999, Casey:2014}. The dust attenuation can be additionally characterized by observing specific features in the SEDs of galaxies such as measuring the Balmer decrement. \ch{Since the dust attenuation parameters are often correlated with each other and other galaxy properties, Bayesian models need to account for this while fitting observations \citep{2022ApJ...932...54N}.} 

The infrared portion of the SED probes dust re-emission and provides crucial constraints on the total dust content. IR SED models typically parameterize the dust emission using templates or modified blackbody functions. The simplest approaches use single-parameter IR templates \citep{2002ApJ...576..159D}, while more sophisticated models can have 4-6 parameters describing various dust temperature components \citep{2004ApJS..152..211Z, 2007ApJ...657..810D, 2015ApJ...806..110D}. For SEDs where the total energy of this process can be conserved (i.e. not resolved regions within galaxies, interacting systems etc.), modern SED fitting approaches fit the UV-optical and IR portions simultaneously using `energy balance' techniques where the total IR luminosity from dust emission is constrained to match the energy of the attenuated light at shorter wavelengths \citep{2005MNRAS.360.1413B, 2015ApJ...806..110D, Salim:2018, 2023MNRAS.525.5720J}. 
\ch{The assumption of energy balance has been challenged in recent times due to observations of spatial decorrelations \citep{2022A&A...665A.137S} between the dust (e.g., cold dust emission seen with ALMA) and star formation (measured through UV emission), violating the underlying assumption of thermal equilibrium\footnote{Energy balance can also be violated by up to a factor of 3 due to the fact that the star-dust geometry in the UV-NIR is not isotropic, in contrast to the FIR emission \citep{2023ApJ...957....7F}.}. While this debate has not been settled at the time of writing, recent work indicates that the assumption of energy balance may hold at high-z \citep{2023MNRAS.525.1535H}, though care needs to be taken in fitting the both inhomogeneous dust attenuation and FIR emission from galaxies using traditional SED fitting methods \citep[][Jespersen et al. \textit{(submitted.)}]{2020A&A...634A..57D, 2024arXiv240919054H}.}
Far-IR and sub-mm telescopes like ALMA and NOEMA have been crucial for our current understanding of the total dust content of galaxies and their properties like dust mass, temperature, and grain-size distributions \citep{2020RSOS....700556H}. Surveys like ALPINE \citep{2020A&A...643A...2B} and REBELS \citep{2022ApJ...931..160B} have been used to estimate the fraction of obscured star formation at high redshifts and study the processes responsible for dust production, growth and destruction \citep{2024MNRAS.528.2407P}. 


The launch of JWST has provided several improvements in our ability to model and constrain dust. The combination of NIRCam and MIRI enables studies of both attenuation and emission at sensitivities that allow us to discriminate between low-z dusty galaxies, high-z galaxies with a Lyman-break, and AGN-dominated galaxies. MIRI allows for detailed spectroscopic measurements of dust emission. Coupled with ALMA observations, these data can be used to derive better constraints on dust composition and grain size distribution, leading to a fuller understanding on the processes responsible for dust formation, evolution and destruction \citep{2023MNRAS.526.2196M, 2023ApJ...944L..14W, 2024A&A...690A.348C, 2023ApJ...944L..23D, 2024arXiv240808962C}.

\subsection{Active Galactic Nucleus (AGN) properties}

The nuclear emission arises from a large number of physical processes that shape different parts of the electromagnetic spectrum, from the low-frequency radio to X-ray and Gamma-rays (see section \ref{sec:agn}). Therefore, the presence of AGN can severely increase the complexity of the parameter space required to model the SEDs of galaxies. To avoid the computational limitations that come with such complex parameter space, many SED fitting codes that are tailored to infer galaxy properties do not include detailed AGN models. 

Nonetheless, modeling the overall SED of the AGN is crucial for two reasons. First, it is crucial to disentangle the nuclear AGN contribution from the galaxy SED, to better infer galaxy properties since AGN emission can dominate over stellar light in certain wavelength regimes and potentially bias estimates of star formation rates and stellar masses if not properly accounted for. Second, for AGN and black hole studies it is crucial to capitalize on the increasing number of multiwavelength surveys and accurately model the AGN emission, so that physical properties of the AGN itself can be inferred, such as black hole, accretion properties, and the nuclear structure.

Additionally, AGN can heat host galaxy dust, contributing to the far-infrared emission traditionally associated with star formation. AGN typically show stronger and more rapid variability than stellar processes, particularly in the UV/Optical, which makes multi-epoch observations useful for isolating the AGN contribution to the UV SED. Spatially resolved observations are also useful for isolating emission from the AGN, which is generally confined to the central regions of its host galaxy. 

Modern AGN SED fitting codes like AGNfitter \citep{2016ApJ...833...98C, 2024A&A...688A..46M}, X-CIGALE \citep{2020MNRAS.491..740Y}, Prospector \citep{2021ApJS..254...22J} and BEAGLE-AGN \citep{2024MNRAS.527.7217V} aim to decompose galaxy SEDs into AGN and host galaxy components through Bayesian inference techniques. These codes typically constrain key physical parameters including the AGN's accretion disk luminosity and temperature, dusty torus properties (optical depth, viewing angle, covering factor), and the relative contribution of AGN to the total galaxy luminosity. The fitting process must account for several challenges, including degeneracies between AGN, star formation and dust emission, AGN variability in multi-epoch data, and the need for broad wavelength coverage from X-ray to radio for optimal constraints.

\section{Modeling \ch{and fitting} a galaxy SED}
\label{sec:fitting}

Armed with the different pieces of what make up a galaxy's SED, we now need to (i) put them together to create an SED given a set of model parameters ($\Theta$), (ii) evaluate the likelihood of this point in parameter space by comparing the modeled SED to a set of observations, and (iii) do this enough times to chart out the likelihood space $\mathcal{L}(\Theta)$, which can then be used to evaluate the credible region of parameter space (also called the posterior distribution) conditioned on the observations. Most modern SED fitting codes use a Bayesian framework for this purpose (for more on Bayesian inference, see the review by \citealt{2022ARA&A..60..363N} and the primer by \citealt{2023arXiv230204703E}).

\subsection{\ch{How did we get here?}}

\ch{Since Beatrice Tinsley's early work \citep{1976ApJ...203...52T} in quantifying the evolving features of stellar populations, the field of SED fitting has come a long way. While early single-band observations necessitated the use of k-corrections and mass-to-light ratios to convert the observed luminosity into a stellar mass, multi-band observations helped better constrain distances to galaxies and incorporate the effects of varying stellar population ages in estimating galaxy masses. As understanding of star formation processes improved, tracers such as UV-emission were used as direct indicators of ongoing star formation. The introduction of rest-frame color-color diagrams like the UVJ diagram led to significant advancements, allowing astronomers to effectively separate star forming and quiescent populations, and providing insights into the dust content of galaxies.}

\ch{With the advent of more sophisticated photometric surveys, researchers could use few-band photometry to fit models that simultaneously vary the star formation history, metallicity and dust attenuation, leading to the creation of modern SED fitting as we know it. \citet{1998AJ....115.1329S} is perhaps the earliest example of this, leveraging the information contained in spectral features across different wavelengths. This was followed by the rapid development of full spectral fitting in the SDSS era, where algorithms like MOPED \citep{2000MNRAS.317..965H} and VESPA \citep{2007MNRAS.381.1252T} were developed to maximize the amount of information inferred from galaxy spectra. This also allowed modelers to begin to relax simplifying assumptions made about star formation histories and dust attenuation, though these would propagate much later to photometric codes. FIR measurements and pan-chromatic observations of galaxies led to the incorporation of dust and AGN emission models in SED fitting codes, but at this point codes were getting difficult to use due to the highly multidimensional parameter space they needed to sample in order to calculate the best-fitting models given a set of observations. Bayesian methods like Markov-chain Monte Carlo (MCMC) techniques were the solution of choice, with most modern SED fitting codes using some form of Bayesian optimization to sample high-dimensional parameter spaces and calculate posteriors on parameters of interest. Going forward, as observations increase in data quality and upcoming facilities set to measure SEDs for $\mathcal{O}(10^8)$ galaxies or more, there is an increasing trend toward using machine-learning based techniques to emulate spectral synthesis, sample high dimensional spaces, and even replace the likelihood computation entirely. Inferring the physical properties of galaxies from their multiwavelength SEDs remains one of the primary ways of probing the properties and evolution of galaxy populations over time, and forms the basis for testing our theories of how galaxies form and evolve.}

\subsection{What can the data constrain?}

The constraining power of SED fitting depends on both the information content of the observations and our ability to model the underlying sources of emission reliably. Different types of data provide varying constraints on different physical parameters, and understanding these relationships is crucial for robust inference.
Broadband photometry, for example, contains significantly less information about narrow spectral features compared to targeted spectroscopy. However, slit spectroscopy might not capture light from the outskirts of a galaxy with a larger angular extent on the sky. The number and wavelength coverage of photometric bands determines which physical properties are accessible through a given set of observations. For example, rest UV-to-NIR photometry can be used to constrain the redshift, stellar mass, SFR, and star formation history, but might not be able to break degeneracies with dust and metallicity, leading to uncertainties in the estimated properties. Adding far-IR photometry helps better constrain the dust temperature and mass, breaking the degeneracy between dust and SFH and giving better constraints on obscured star formation. Further adding $R\sim 100-1000$ spectroscopic data provides estimates of emission and absorption line strengths, constraining metallicity and fully resolving the age-dust-metallicity degeneracy. An additional factor over the coverage is also the signal-to-noise of the observations, which determine the extent to which the posteriors are dominated by the prior instead of the likelihood. It is important to understand the limitations of the available data given the scientific question, and to tailor the SED fitting approach accordingly. Attempting to constrain too many parameters with too few or uninformative data points can lead to degeneracies and unreliable results.

\subsection{Noise and nuisance parameters}

Galaxy SED fitting must account for various nuisance parameters that affect observations but are not of direct scientific interest. These include aleatoric uncertainties (inherent to the data) like photon noise in flux measurements, sky background fluctuations, detector read noise, and source confusion effects. Sometimes, a careful reduction of the raw data accounting for the source of these different types of noise and their resulting contributions can lead to observations that are significantly deeper than the default application of standard reduction pipelines. In addition, uncertainties in wavelength and/or flux calibration often require the introduction of additional parameters (such as a 
polynomial correction term) in order to robustly constrain the parameters of interest\footnote{Without the additional parameters, the fitting code will still estimate a credible region, but often with artificially narrow posteriors and possible biases.}. Marginalization over these parameters can be performed either analytically (when possible) or numerically through sampling. The treatment of these nuisance parameters can significantly impact the derived uncertainties on physical parameters of interest. 

In addition to the uncertainties on individual galaxies, the entire galaxy catalog can also be used for ensemble level corrections including overall uncertainty calibration (by studying the distribution of residuals after fitting to determine whether the reported uncertainties are over- or under-estimated), and for zeropoint corrections\footnote{Another term, the K-correction \citep{2002astro.ph.10394H}, was used to account for the effect of redshift on the observed colors of galaxies. Because the spectral energy distribution of a galaxy is redshifted when observed at higher redshifts, its colors will appear different than they would if the galaxy were observed at rest. The k-correction is a wavelength-dependent correction factor that converts the observed colors of a galaxy to the colors that would be observed if the galaxy were at a reference redshift (usually z=0). Since we account for the redshift as a free parameter in modern SED fitting tools, this procedure is generally not needed any more. \ch{It is, however, required when comparing populations of galaxies across a non-discrete redshift range, in order to make inferences on the evolution of galaxy populations with redshift in a fair manner.}}. Zeropoints are the magnitudes of a star that would produce a flux of one count per second in a given photometric band. They are used to convert instrumental magnitudes (relative to an arbitrary reference point) to calibrated, absolute magnitudes. Zeropoints were typically determined by observing standard stars with known magnitudes and comparing their instrumental magnitudes to their true values, but often contain second-order corrections due to instrumental effects. In modern galaxy catalogs, they are often determined empirically with codes like EAZY \citep{2006AJ....132..926C, 2007AJ....133..734B, 2008ApJ...686.1503B, 2014MNRAS.441.2891M} by deriving a correction factor that minimizes the differences between photometric and spectroscopic redshifts.

The second source of uncertainties (also called epistemic uncertainties; \citealt{2022ApJ...935..138G}) come from modeling limitations and choices. In traditional SED fitting, this can be factors like the choice of SFH or dust model (e.g., a parametric model might not be able to access portions of observable space that a non-parametric model can, or a simple dust law might not be able to fit both the rest-optical and UV due to differential attenuation of young stars), but also overall choices like SPS model. In a machine learning approach, these can be factors like the amount of training data and how representative it is of the actual observations being analyzed. 

\subsection{Likelihoods}

Bayesian inference techniques dominate the current landscape of SED fitting. These methods involve defining prior probability distributions for the model parameters and using Bayes' theorem to compute the posterior probability distribution given the observed data, written as $p(\Theta | F_{obs}) = p(F_{obs} | \Theta) p(\Theta) / (\int p(F_{obs} | \Theta) p(\Theta) d\Theta)$, written colloquially as \textit{posteriors = likelihood $\times$ priors / evidence} \citep{2016MNRAS.462.1415C}. Bayesian methods can provide more robust parameter estimates and uncertainties, and they can be more efficient than grid-based methods by focusing the computational resources on the regions of parameter space that are most consistent with the data.

The likelihood function $\mathcal{L}(F_{obs} | \Theta) \equiv p(F_{obs} | \Theta)$ quantifies the probability of the model parameters $\Theta$ given the observed data $F_{obs}$. For galaxy fitting, the likelihood typically assumes the form\footnote{While the Gaussian likelihood remains the most commonly used (by far), likelihoods based on distributions like Student-t are more robust to outliers and might be better suited for specific scientific questions. Other likelihoods are based on distance metrics like the Wasserstein distance or information-theoretic like the KL divergence.}:
\begin{equation}
    \mathcal{L}(\Theta | F_{obs}) = \prod_i \frac{1}{\sqrt{2 \pi \sigma_i^2}} \exp \left( -\frac{(F_{obs}^i - F_{model}^i(\Theta))^2}{2\sigma_i^2} \right)
\end{equation}
where $F_{obs}^i$ represents observed fluxes or intensities at a given wavelength or in a given filter, $F_{model}^i(\Theta)$ the model predictions, and $\sigma_i$ the measurement uncertainties (assuming uncorrelated errors). For correlated errors, this is replaced by a full matrix based likelihood $\chi = X^T \Sigma^{-1} X $ where $X = (F_{obs}^i - F_{model}^i(\Theta))^2$. For resolved fitting, the likelihood must account for pixel-to-pixel (or spaxel-to-spaxel) correlations and point spread function (PSF) effects. In cases where an analytic likelihood is not tractable, modern machine learning methods allow for implicit likelihood inference (also called likelihood-free inference or simulation-based inference) where a forward model or set of labeled data with uncertainties can be used to train a kind of neural network called a normalizing flow to learn an implicit likelihood function that can then be used for inference \citep{2023ApJS..268....7W, 2024OJAp....7E..54H}. 

\subsection{Priors}

Prior distributions $P(\Theta)$ in Bayesian SED fitting encode our physical understanding of galaxy properties before considering the data\footnote{While Bayesian analysis recommends that the prior be as uninformative as possible, it is important to remember that this ignorance pertains to the parameter value, not the parameter distribution itself. Any choice of distribution constitutes perfect knowledge of the prior, but the goal should be to capture ignorance about the parameter value \citep{1946RSPSA.186..453J}.}. Common priors include physical constraints (such as non-negative ages or star formation rates), causality (e.g., stars can not have ages older than the universe), or conservation (e.g. energy conservation between UV/optical and IR while modeling dust). They can also incorporate population-level information such as galaxy scaling relations, mass and luminosity functions, or environmental dependencies. It is important to acknowledge that implicit modeling choices, such as the parametrization of SFH or dust law, with `flat' priors on the parameters, also impose strong priors on the resulting distribution of SFR as a function of cosmic time, or attenuation as a function of wavelength, respectively. The choice of priors becomes particularly important when dealing with degeneracies in the parameter space or low signal-to-noise observations. Modern approaches often employ weakly informative priors that constrain parameters to physically reasonable ranges without strongly influencing the posterior.
Priors can sometimes include free parameters that can be learned at the ensemble level before being applied to individual galaxies \citep{2021PhRvD.104h3008M}, although one should be careful to avoid `double dipping', where the same dataset is used to both estimate the prior hyperparameters and subsequent fitting using those parameters.

\subsection{Samplers}

The complexity of galaxy fitting necessitates sophisticated algorithms to explore the posterior distribution $P(\Theta | F_{obs}) \propto \mathcal{L}(\Theta | F_{obs}) P(\Theta)$. A straightforward approach to SED fitting is to generate a large grid of models spanning a range of parameter values and to find the best-fitting model by exhaustively searching the grid. 
While early approaches to SED fitting used grid-based quadratic minimization, it became impractical as the dimension of the parameter space ($\Theta$) grew larger as SFH and dust models grew more complex. Eventually, Markov-chain Monte Carlo (MCMC) based methods were widely adopted for efficiently sampling the parameter space \citep[for a detailed description, see][]{2011ApJ...737...47A, 2013PASP..125..306F, 2019arXiv190912313S}. MCMC methods spend a lot of time sampling the posterior maxima, and were revolutionary for estimating parameter uncertainties and covariances in comparison to the traditional `best-fit' or median estimates for the physical properties of galaxies. They also provide significant increases in speed compared to grid-based methods, since they do not need to spend time sampling low-probability regions of parameter space. However, they MCMC based methods tend to fail in situations where the likelihood surface has pathological features or is heavily multimodal. This can be addressed to some extent with multiple walkers and affine methods, but the next development came in the form of Nested Sampling \citep{2009MNRAS.398.1601F, 2020MNRAS.493.3132S}, which is focused on calculating the Bayesian evidence instead of the posterior peak. It does so by estimating isolikelihood contours, which are then used to integrate over the prior volume to estimate the evidence. This has the advantage of much more robust coverage over the posterior space, and ability to accurately capture large degeneracies and multimodal posteriors. 

Current SED fitting codes use a mix of grid based methods, MCMC, nested sampling, and machine-learning based sampling. Modern grid-based approaches can be often made tractable by (i) using a grid based on cosmological simulations \citep{2012MNRAS.421.2002P} that span only the physically self-consistent part of parameter space, (ii) using samples from correlated priors instead of a uniform grid that more efficiently cover high-z parameter spaces, and (iii) training modern probabilistic neural networks like normalizing flows to learn the mapping between SEDs and parameter space. Some codes also use amortized inference (sometimes referred to as `brute-force Bayesian' methods), which involves precomputing a large set of synthetic observations corresponding to different parts of parameter space (either sampled uniformly or randomly from a prior distribution) and then using these samples while fitting to significantly speed up inference. This is particularly advantageous for fitting large catalogs of galaxies, by offloading the computationally intensive portion of a likelihood call (i.e., computing a spectrum corresponding to a point in parameter space) to a single step in advance of the actual fitting, thereby reducing the amount of time it takes to fit any single galaxy. An alternative to amortized computation is to train a neural emulator \citep{2023ApJ...954..132M}, that can learn the effective mapping between the parameter space and galaxy SEDs.
The choice of sampler depends on factors including parameter space dimensionality, likelihood expense, and the need for evidence calculations. Modern implementations often also employ parallel computing to accelerate convergence.

\subsection{From brute force to deep learning: the applications of ML methods in SED modeling}

In keeping with the rapid increases in data quality and volume, approaches to SED fitting have also evolved over time, from simple grid-based or template-based methods to more sophisticated statistical techniques and machine learning algorithms. 
\ch{One such example includes the use of unsupervised clustering and dimensionality reduction algorithms (often grouped under the term `manifold learning') to organize galaxy SEDs or colors by similarity, using techniques like self-organizing maps (SOMs; \citealt{1988soam.book.....K}, applications in \citealt{2012MNRAS.419.2633G, 2014MNRAS.438.3409C, 2015ApJ...813...53M, 2024AJ....167..261L}), dimensionality reduction based on a distance metric  (t-SNE,  UMAP, other distance metrics; e.g. the Sequencer, \citealt{2021ApJ...916...91B}), or latent spaces using neural architectures like variational autoencoders (VAEs; \citealt{kingma2013auto},  applications in \citealt{2020AJ....160...45P, 2023arXiv231216687G}) (see review by \citealt{2024A&C....4800851F} for more details). Once the data manifold has been constructed, predictions of physical properties can be obtained by simply projecting any observation into this manifold and sampling the properties of other labeled galaxies close to it. These have the advantages of finding data-driven correlations in the observations, propagating sparse labels to larger datasets, and often being computationally inexpensive after the training step, and are being used as part of large upcoming surveys \citep{2024ApJ...967...60J, 2024A&A...689A.144T, 2024A&A...692A.177J}. Another area where machine learning techniques have been extensively used is determining outliers and interesting galaxies for further studies. In addition to manifold learning, techniques such as isolation forests and active learning or `human-in-the-loop' ML methods \citep{2017MNRAS.465.4530B, 2021A&C....3600481L} have been very successful at finding and labeling niche classes of galaxies.} 

In the last few years, deep learning techniques have emerged as a promising approach to SED fitting. Deep learning algorithms, such as convolutional neural networks (CNNs) and generative adversarial networks (GANs), can learn to extract features from SEDs and to directly learn the relationship between spectral features and intrinsic properties \citep{2019MNRAS.490.5503L, 2020MNRAS.493.4808S, 2021ApJ...916...43G}. These methods have the potential to be much faster and more flexible than traditional SED fitting techniques, but they require large training datasets and careful validation to ensure their reliability.

Some recent applications include 
predicting SFHs with CNNS \citep{2019MNRAS.490.5503L}, 
predicting the recent SFH using approximate Bayesian computation \citep{2020A&A...635A.136A}, 
predicting stellar mass, SFR, and dust luminosity \citep{2020MNRAS.493.4808S}, 
using unsupervised labels and tree-based methods to estimate stellar mass, SFR and redshift \citep{2023MNRAS.520..305H}, 
predicting stellar mass, SFR, and metallicity \citep{2024A&A...688A..33Z}, 
predicting SFR in large galaxies using ML \citep{2024arXiv241006736R}, 
predicting spectral features, HI mass fraction, and SFHs from galaxy imaging instead of SEDs \citep{2020ApJ...900..142W, 2020arXiv200912318W, 2024ApJ...967..152A}, and
using normalizing flows to constrain cosmology and feedback from mass functions and color distributions \citep{2024arXiv241113960L}. 
More recently, simulation based inference methods have emerged as a powerful new approach for performing SED fitting, allowing for amortized posteriors which greatly speed up the process of obtaining robust posterior parameter distributions \citep{2024arXiv240509720C, 2024A&A...686A.133B|, 2024A&A...689A..58I}. \ch{The development of techniques like SBI are also enabling the rise of fully `forward-modeled' approaches to inferring population-level physical properties from observations (e.g., \citealt{2024AJ....167...16L, 2024ApJ...975..145T, 2024ApJ...961...53I}), in contrast to traditional SED fitting approaches that need to be applied on a galaxy-by-galaxy basis to first reconstruct physical properties and then use the posteriors to build scaling relations.}
This is by no means an exhaustive list, and innovative astrostatistical and machine learning methods are being applied to the problem of inferring the physical properties of galaxies with every passing week. 

\subsection{Diagnostics}

Given the complexity of a galaxy's SED, the reliability of SED fitting depends critically on careful validation and diagnostic checks. Modern approaches use multiple complementary diagnostics to assess both the quality of fits to individual galaxies and the overall performance of the fitting methodology.

\subsubsection{What does it mean to `fit' an SED?}

\begin{figure*}
    \centering
    \includegraphics[width=0.9\linewidth]{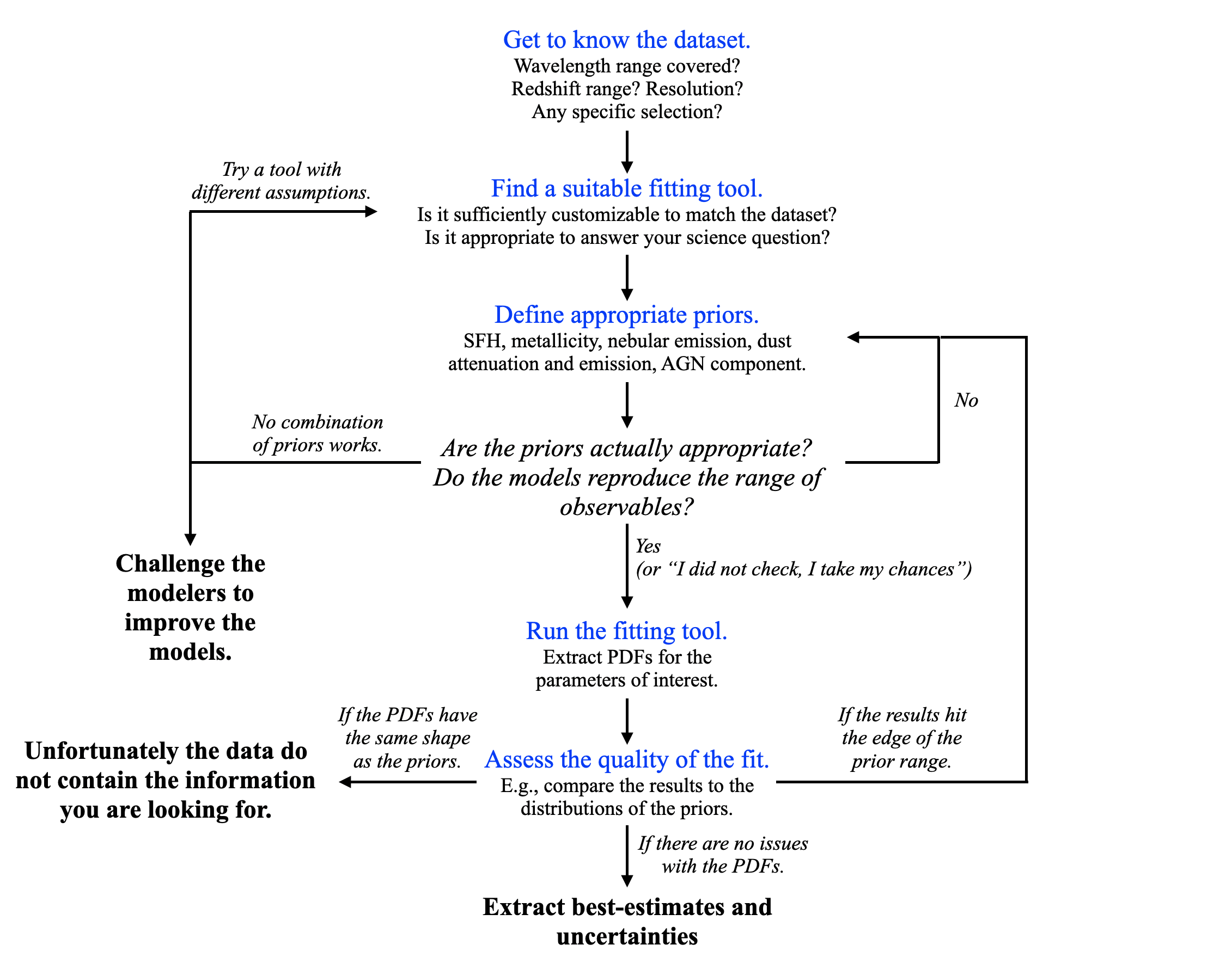}
    \caption{Flowchart outlining the basic SED fitting workflow, which requires understanding the dataset, setting the priors, testing modeling assumptions, running the fits with the selected tool, and evaluating whether the results are robust and sensible, reproduced from \cite{2023ApJ...944..141P}.}
    \label{fig:sedfittingchart}
\end{figure*}

At its core, SED fitting involves finding the range of parameter space that best reproduces the observed SED of a galaxy. This is typically done by maximizing a likelihood based on the similarity between the observed and model SEDs, using metrics such as the $\chi^2$ statistic. However, it is important to recognize that a `best fit' model is not necessarily a robust (or even accurate) description of the underlying SED. In other words, \textit{a fit will always return a result, but the user needs to understand when the result is meaningful}. 

A common pitfall in SED fitting is over-interpreting the results, especially when the data are of poor quality or the models are not well-suited to the object being studied. For example, fitting a complex star formation history to a galaxy with only a few broad-band photometric measurements is likely to yield highly uncertain and potentially misleading results. It is crucial to always assess the posteriors from fitting and to consider the uncertainties and degeneracies in the model parameters, generally following the workflow described in Figure \ref{fig:sedfittingchart}.

\subsubsection{Goodness-of-fit, Residuals, and Likelihood distributions}

A basic diagnostic of a fit is the residual between an optimal model and the observed SED, typically quantified through the $\chi=(F_{\Theta} - F_{obs})/\sigma_{obs}$ statistic or likelihood values. However, $\chi^2$ values can be misleading due to template incompleteness, parameter degeneracies, and uncertainties in both models and data. Instead, examining the full posterior distribution and the structure of residuals across wavelengths provides deeper insight. Systematic patterns in residuals can reveal inadequacies in the assumed dust law, stellar population models, or other components. The distribution of likelihood values across the sample should also be examined - highly skewed or multimodal distributions may indicate problems with the model assumptions or degeneracies between parameters.

\subsubsection{Bayesian Evidence and Model Comparison}

The Bayesian evidence (also called the marginal likelihood or prior predictive density) provides a principled way to compare different model choices, from simple questions like the preferred dust law or SFH parametization to more fundamental choices about stellar population models \citep{2021MNRAS.502.3993L, 2023ApJS..269...39H}. The evidence naturally penalizes more complex models unless they provide significantly better fits to the data. However, evidence values must be interpreted carefully, considering both the priors and the effective complexity of different models. Model comparison should examine not just overall evidence values but also how well different models reproduce key observables like colors, line ratios, and derived physical quantities. Cross-validation approaches, where models are fit to one subset of the data and validated on another, can help assess predictive performance and guard against overfitting.

\subsubsection{Color-Color Diagrams and Scaling Relations}

Color-color diagrams and scaling relations between derived physical properties such as the SFR-M$_*$ or mass-metallicity plane provide important external validation checks. The distribution of fitted galaxies in color space should match known relations, accounting for selection effects. Similarly, derived quantities like stellar masses and star formation rates should follow established scaling relations like the star-forming main sequence. Significant deviations may indicate problems with the fitting methodology. These spaces are also useful for finding attractor solutions, where stripes or odd clustering can indicate that a model is being favored during the fitting, either due to a parameter degeneracy or overfitting to noise.

\subsubsection{Posterior predictive checks}

Posterior predictive checks compare simulated data generated from the fitted model to the actual observations that were not included in the original fits. These checks can reveal subtle mismatch between model assumptions and data that may not be apparent from residuals alone. 

\subsubsection{Comparing Against Synthetic Observations}

Testing SED fitting codes on synthetic observations from cosmological simulations, where the true physical properties are known, provides a powerful validation tool \citep[e.g.,][]{2012MNRAS.422.3285P, 2015MNRAS.446.1512H, 2015MNRAS.454.2381G, 2017ApJ...838..127I, 2020ApJ...904...33L, 2024ApJ...975..220F}. Such tests can quantify systematic uncertainties and biases in derived properties. However, care must be taken since simulations may not capture all relevant physics. Ideally, codes should be tested on multiple simulation suites with different physical models.

\subsubsection{Inter-code comparisons}

Comparing posteriors estimated using different codes applied to the same observations can highlight the effects of differing modeling assumptions and priors. Studies like \cite{2013ApJ...775...93D, 2023ApJ...944..141P} compare a range of photo-z and SED fitting codes to quantify the inter-code variability and the associated epistemic uncertainties on different physical properties. 

\subsubsection{Self-consistency checks}

A final class of diagnostic checks involves self-consistency with the evolution of galaxies over time. For example, computing the sum of the SFHs for an ensemble of galaxies corrected for selection effects should yield the cosmic star formation rate density (SFRD) up to that epoch, modulo a correction term for mergers \citep{2019ApJ...873...44C, 2019ApJ...876....3L}. Similarly, tracing galaxies backwards in time along their SFHs should yield mass functions and scaling relations consistent with those observed directly at earlier epochs \citep{2018ApJ...866..120I, 2019MNRAS.482.1557S}. 

The combination of these diagnostics provides a comprehensive assessment of SED fitting results. No single diagnostic is definitive - apparent problems in one metric may be resolved by examining others. Regular application of these tests is crucial as data quality improves and models become more sophisticated.

Ultimately, the choice of SED fitting approach depends on the specific goals and constraints of the problem at hand. Brute force methods may be sufficient for simple problems with small parameter spaces, while Bayesian methods may be more appropriate for problems with complex likelihood functions and prior information. Deep learning methods may be advantageous for problems with large datasets and where fast, automated processing is required.

Regardless of the approach used, it is important to always critically assess the results of SED fitting and to be mindful of the limitations and assumptions involved. By understanding the philosophy behind SED fitting, astronomers can make more informed decisions about how to best extract physical insights from their observations.

\section{Conclusions}

The study of galaxy spectral energy distributions is one of the most powerful tools in modern astrophysics for understanding how galaxies form and evolve across cosmic time. A galaxy's SED encodes the rich history of present and past star formation activity, chemical enrichment, dust production, and supermassive black hole activity. Through increasingly sophisticated observational techniques and theoretical modeling, astronomers can decode these spectral fingerprints to constrain the physical processes driving galaxy evolution.

The remarkable progress in SED analysis over the past few decades has been driven by complementary advances on multiple fronts. Theoretically, our models now incorporate complex physics ranging from stellar evolution and radiative transfer to the effects of dust and AGN activity. Observationally, revolutionary facilities have expanded our view across the electromagnetic spectrum, from X-rays to radio waves, while large surveys have built statistical samples of galaxies across cosmic time. Computationally, new techniques in Bayesian analysis and machine learning have transformed our ability to extract physical insights from these rich datasets.

Yet significant challenges and opportunities remain. The degeneracies between different physical processes that can produce similar spectral features continue to complicate the interpretation of SEDs. Our understanding of dust physics, AGN feedback, and the impacts of galaxy environment remains incomplete. The high-redshift universe, where we observe galaxies in their infancy, presents particular challenges for SED fitting due to limited wavelength coverage and signal-to-noise.

The next decade promises to be transformative for the field. New facilities like JWST are already providing unprecedented views of galaxy evolution in the early universe. Future observatories will gather spectral data for hundreds of millions of galaxies, enabling population studies of unprecedented statistical power. These observations, combined with advances in theoretical modeling and computational techniques, will help answer fundamental questions about how galaxies acquire their mass, form their stars, and evolve into the diverse population we observe today.

As our understanding of galaxy SEDs continues to deepen, this field remains at the forefront of addressing some of astronomy's most profound questions: How did the first galaxies form? What drives the diversity of galaxy properties we observe? How do galaxies evolve from active star-forming systems to quiescent objects? Through the cosmic stories written in their spectra, galaxies continue to reveal their secrets, one wavelength at a time.

\begin{ack}[Acknowledgments]

KI would like to thank \ch{Eric Gawiser, Charlotte Olsen, Greg Bryan, Steve Finkelstein, Marcin Sawicki, Gael Noirot and Guillaume Desprez} for feedback and support during the writing of this manuscript, and Sandy Faber, Harry Ferguson, Viviana Acquaviva, Casey Papovich, Bob Abraham, Lamiya Mowla, Rachel Somerville, Adam Carnall and Joel Leja for discussions on SED fitting (both technical and philosophical), and Noopur Gosavi for infinite moral support. We would also like to thank all the people who develop and maintain public SED fitting codes. 
Support for KI was provided by NASA through the NASA Hubble Fellowship grant HST-HF2-51508 awarded by the Space Telescope Science Institute, which is operated by the Association of Universities for Research in Astronomy, Inc., for NASA, under contract NAS5-26555.
\end{ack}

\section{SED fitting codes and resources}

In this section, we have tried to compile a list of SED fitting methods currently described in the literature\footnote{While we have tried to be complete, the vast literature makes it difficult to capture every new method, which requires a living list similar to the one at \href{http://www.sedfitting.org/Models.html}{sedfitting.org}.}. 
Where possible, we have provided links to repositories containing the code for public SED fitting packages. 
This list will be available at \href{https://sites.google.com/view/sed-fitting-forum/sed-fitting-codes}{on the pan-survey SED forum website}, and will be updated with new SED fitting codes as they become available.

\begin{enumerate}
    \item (quadratic minimization) \citep{1972A&A....20..361F} 
    \item SEDfit \citep{1998AJ....115.1329S, 2012PASP..124.1208S}
    \item LePhare \citep{1999MNRAS.310..540A, 2006A&A...457..841I} - code available at \url{https://gitlab.lam.fr/Galaxies/LEPHARE}
    \item MOPED \citep{2000MNRAS.317..965H} - code available at \url{https://github.com/heatherprince/cosmoped}
    \item zphot \citep{2000AJ....120.2206F, 2019MNRAS.490.3309M}
    \item Hyper-z \citep{2000A&A...363..476B} - code available at \url{http://www.bo.astro.it/~micol/Hyperz/old_public_v1/index.html}
    \item FITSED \citep{2001ApJ...559..620P, 2015ApJ...799..183S}
    \item SINOPSIS \citep{2001ApJ...550..195P, 2014A&A...566A..32F} - code available at \url{https://www.irya.unam.mx/gente/j.fritz/JFhp/SINOPSIS.html} 
    \item pPXF \citep{2004PASP..116..138C, 2023MNRAS.526.3273C} - code available at \url{https://pypi.org/project/ppxf/}
    \item CHORIZOS \citep{2004PASP..116..859M}
    \item Starlight \citep{2005MNRAS.358..363C} - code available at \url{http://www.starlight.ufsc.br/}
    \item Stecmap \citep{2006MNRAS.365...46O} - code available at \url{https://github.com/pocvirk/STECKMAP }
    \item VESPA \citep{2007MNRAS.381.1252T} - code available at \url{http://www-wfau.roe.ac.uk/vespa/ }
    \item eazy (and eazypy) \citep{2008ApJ...686.1503B, 2023zndo...8268031B} \url{https://eazy-py.readthedocs.io}
    \item MAGPHYS \citep{2008MNRAS.388.1595D, 2019ApJ...882...61B} - code available at \url{http://www.iap.fr/magphys/}
    \item GOSSIP \citep{2008ASPC..394..642F} - code available at \url{https://ascl.net/1210.003}
    \item FAST \citep{2009ApJ...700..221K, 2017MNRAS.465.3390A} - code available at \url{https://github.com/jamesaird/FAST}
    \item ULySS \citep{2009A&A...501.1269K} - code available at \url{http://ulyss.univ-lyon1.fr/}
    \item GalMC / SpeedyMC \citep{2011ApJ...737...47A, 2015ApJ...804....8A}
    \item (semi-analytic SFHs) \citep{2012MNRAS.421.2002P, 2024arXiv240807749G}
    \item (manifold learning; SOMs) \citep{2012MNRAS.419.2633G, 2015ApJ...813...53M, 2019ApJ...881L..14H} - code available at \url{https://ascl.net/1404.014}
    \item CMCIRSED \citep{2012MNRAS.425.3094C} - code available at \url{https://www.as.utexas.edu/~cmcasey/cmcirsed.html}
    \item iSEDfit \citep{2013ApJ...767...50M} - code available at \url{https://github.com/moustakas/iSEDfit}
    \item SED3FIT \citep{2013A&A...551A.100B} \url{https://cosmos.astro.caltech.edu/page/other-tools}
    \item BayeSED \citep{2014ApJS..215....2H, 2023ApJS..269...39H} \url{https://github.com/hanyk/BayeSED3/}
    \item SEABASs \citep{2014MNRAS.438..494R, 2018ApJ...857...31T}
    \item IZI \citep{2015ApJ...798...99B} - code available at \url{https://users.obs.carnegiescience.edu/gblancm/izi/}
    \item Beagle / Beagle-AGN \citep{2016MNRAS.462.1415C, 2024MNRAS.527.7217V} - code available at \url{https://www.iap.fr/beagle/}
    \item pipe3d / fit3d \citep{2016RMxAA..52...21S, 2022NewA...9701895L} - code available at \url{https://gitlab.com/pipe3d/pyPipe3D}
    \item AGNFitter / AGNFitter-RX \citep{2016ApJ...833...98C, 2024A&A...688A..46M} - code available at \url{https://github.com/GabrielaCR/AGNfitter}
    \item dense basis \citep{2017ApJ...838..127I, 2019ApJ...879..116I} - code available at \url{https://dense-basis.readthedocs.io} 
    \item Prospector \citep{2017ApJ...837..170L, 2021ApJS..254...22J} - code available at \url{https://prospect.readthedocs.io}
    \item Lightning \citep{2017ApJ...851...10E, 2023ApJS..266...39D} - code available at \url{https://github.com/rafaeleufrasio/lightning}
    \item Firefly \citep{2017MNRAS.472.4297W, 2022MNRAS.513.5988N} - code available at \url{https://github.com/FireflySpectra/firefly_release}
    \item FADO \citep{2017A&A...603A..63G} - code available at \url{http://cdsarc.u-strasbg.fr/viz-bin/qcat?J/A+A/603/A63}
    \item ALF \citep{2018ApJ...854..139C, 2023ApJ...947...13O} - code available at \url{https://github.com/cconroy20/alf}
    \item BAGPIPES \citep{2018MNRAS.480.4379C} - code available at \url{https://bagpipes.readthedocs.io}
    \item Mr-Moose \citep{2018MNRAS.477.4981D} - code available at \url{https://github.com/gdrouart/MrMoose}
    \item CIGALE / X-CIGALE \citep{2019A&A...622A.103B, 2020MNRAS.491..740Y} - code available at \url{https://cigale.lam.fr/}
    \item FortesFit \citep{2019ascl.soft04011R} - code available at \url{https://github.com/vikalibrate/FortesFit}
    \item PEGASE.3 \citep{2019A&A...623A.143F} - code available at \url{https://www2.iap.fr/users/fioc/Pegase/Pegase.3/}
    \item michi2 \citep{2019ApJS..244...40L} - code available at \url{https://github.com/1054/Crab.Toolkit.michi2}
    \item (approximate Bayesian computation) \citep{2020A&A...635A.136A}
    \item ProSpect \citep{2020MNRAS.495..905R} - code available at \url{https://github.com/asgr/ProSpect}; Online SED explorer at \url{https://prospect.icrar.org/}
    \item MCsed \citep{2020ApJ...899....7B} - code available at \url{https://mcsed.readthedocs.io}
    \item pixedfit \citep{2021ApJS..254...15A} - code available at \url{https://pixedfit.readthedocs.io}
    \item Phosphoros \citep[][Paltani et al., in prep.]{2020A&A...644A..31E} - code available at \url{https://phosphoros.readthedocs.io/en/latest/}
    \item Mirkwood \citep{2021ApJ...916...43G} - code available at \url{https://github.com/astrogilda/mirkwood}
    \item IRAGNSEP \citep{2021MNRAS.503.2598B} - code available at \url{https://pypi.org/project/iragnsep/}
    \item Starduster \citep{2022ApJ...930...66Q} - code available at \url{https://github.com/yqiuu/starduster}
    \item (semi-analytic spectral fitting) \citep{2022MNRAS.513.5446Z}
    \item SedFlow/PROVABGS \citep{2022ApJ...938...11H, 2023ApJ...945...16H} - code available at \url{https://github.com/changhoonhahn/SEDflow} and \url{https://github.com/changhoonhahn/provabgs}
    \item parrot.jl (neural emulators) \citep{2023ApJ...954..132M} - code available at \url{https://github.com/elijahmathews/MathewsEtAl2023}
    \item (simulation based inference) \citep{2024A&A...689A..58I}
    \item FastSpecFit (\citealt{2023ascl.soft08005M}, Moustakas et al. \textit{in prep.}) - code available at \url{https://fastspecfit.readthedocs.io/}
    \item SMART \citep{2024MNRAS.531.2304V} - code available at \url{https://github.com/ch-var/SMART}
    \item Starnet (CNN) \citep{2024MNRAS.530.4260W}
    \item pop-cosmos \citep{2024ApJS..274...12A, 2024ApJ...975..145T}
    \item GalaPy \citep{2024A&A...685A.161R} - code available at \url{https://galapy.readthedocs.io}    
    \item SEW \citep{2025arXiv250213696L} - code available at \url{https://pypi.org/project/ppxf-sew/} 
\end{enumerate}

\section{SED fitting systematics and review articles}

\ch{We list some review articles and references here that are relevant to modeling and fitting galaxy SEDs. While the chapter generally strives to be self-complete, the purpose is to provide a broad pedagogical overview of the current state of the field instead of a detailed, in-depth exploration. Additional information can be found in other chapters of this encyclopedia (such as those on stellar evolution and population synthesis, properties of AGN, and the interstellar and intergalactic media), with the references below providing additional resources for further reading.} 

\begin{enumerate}
    \item \citet{2011Ap&SS.331....1W, 2013ARA&A..51..393C}: Reviews on modeling the panchromatic SEDs of galaxies, still an authoritative source a decade later.
    \item Other related reviews: 
    IMF \citep{2003PASP..115..763C}, 
    AGN \citep{1993ARA&A..31..473A, 2000ARA&A..38..521S, 2012ARA&A..50..455F, 2013ARA&A..51..511K, 2014ARA&A..52..529Y, 2015ARA&A..53..365N},
    star formation \citep{1980FCPh....5..287T, 2004ARA&A..42...79B, 2007ARA&A..45..565M, 2012ARA&A..50..531K, 2012ARA&A..50..251I, 2020ARA&A..58..727J, 2023ASPC..534..275O, 2022ARA&A..60..455E, 2023ARA&A..61...65K, 2024A&A...685A..40S}, 
    gas and ISM \citep{1980FCPh....5..287T, 1997ApJ...475..479W, 2005ARA&A..43..861W, 2013ARA&A..51..207B, 2019ARA&A..57..511K, 2020ARA&A..58..157T, 2020RSOS....700556H, 2020ARA&A..58..617O, 2022ARA&A..60..319S}, 
    dust \citep{2020ARA&A..58..529S, 2018NewAR..80....1D, 2014ARA&A..52..373L}, 
    CGM \citep{2017ARA&A..55..389T, 2020ARA&A..58..363P, 2023ARA&A..61..131F}, 
    IGM \citep{1998ARA&A..36..267R, 2016ARA&A..54..313M, 2023ARA&A..61..373F}, 
    theory of galaxy formation \citep{2015ARA&A..53...51S}, and 
    photometric redshifts \citep{2022ARA&A..60..363N}.
    \item Further reading on stellar population synthesis models: \citet{2003MNRAS.344.1000B, 2009ApJ...699..486C, 2010ApJ...712..833C, 2010MNRAS.404.1639V, 2011ApJ...740...13Z, 2012MNRAS.427..127B, 2016ApJ...823..102C, 2017PASA...34...58E, 2023ApJS..264...45F, 2023MNRAS.521.4995B, 2025MNRAS.tmp..163B, 2025arXiv250103133N}
    \item Further reading on photoionization models and emission lines: \citet{1998PASP..110..761F, 2008ApJS..178...20A, 2021ApJ...922..170B, 2022ApJ...927...37J, 2024A&A...691A.173B, 2024arXiv240504598L}
    \item \citet{2023ApJ...944..141P} compares different SED fitting codes applied to the same observed data and quantifies uncertainties due to modeling assumptions. \citet{2024MNRAS.530.4260W} perform a similar comparison using multiple codes applied to synthetic SDSS spectra. Numerous papers \citep{2012MNRAS.422.3285P, 2013ApJ...775...93D, 2015MNRAS.446.1512H, 2015MNRAS.454.2381G, 2017ApJ...838..127I, 2018ApJ...853..131L, 2019ApJS..240....3H, 2020ApJ...904...33L,  2023SCPMA..6629513X, 2024ApJ...975..220F, 2024arXiv241017697R, 2024arXiv241017698B} compare the effects of varying SED modeling assumptions on a range of derived physical properties of interest, either on observational datasets or synthetic data forward-modeled from simulations. 
\end{enumerate}

\seealso{Star Formation; Stellar Initial Mass Function; Evolution and the Final Fates of Massive Stars; Population Synthesis; AGN Across Cosmic Time; Unified Models of AGN; Star formation and stellar mass history of the universe; Cosmic Quenching; Chemical Evolution of Galaxies; Gas in Galaxies; The Interstellar Medium; Intergalactic Medium; First Galaxies; Cosmology with HI}

\bibliographystyle{Harvard}
\bibliography{els-article}

\end{document}